\documentclass[twocolumn,aps,pr,superscriptaddress,preprintnumbers,nofootinbib,10pt]{revtex4-2}
\usepackage{amsmath,amssymb}
\usepackage[dvipdf,dvips]{graphicx}
\usepackage{color}
\usepackage{hyperref}
\usepackage{url}
\usepackage{slashed}
\usepackage{subfigure}
\usepackage[usenames,dvipsnames]{xcolor}
\usepackage{amsmath}
\usepackage{amsfonts}
\usepackage{float} 
\usepackage{amssymb}
\usepackage{epsfig}
\usepackage{graphics}
\usepackage{euscript}
\usepackage{slashed}
\usepackage{epstopdf}
\usepackage[utf8]{inputenc}
\usepackage{bbold}
\usepackage{pifont}
\usepackage{dsfont}
\usepackage{MnSymbol}
\allowdisplaybreaks

\hypersetup{
colorlinks=true,
citecolor=blue,
citebordercolor=red,
linktoc=all,
linkcolor=blue,
urlcolor=blue
}

\usepackage{verbatim}
\usepackage{graphicx}
\usepackage{latexsym}

\graphicspath{{./figs/}}
\newbox{\ORCIDicon}
\sbox{\ORCIDicon}{\large
                  \includegraphics[width=0.8em]{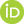}}

\begin{document}

\title{Infrared gluon propagator in the Refined Gribov-Zwanziger scenario\\
 at one-loop order in the Landau gauge}

\author{Gustavo P. de Brito\,\href{https://orcid.org/0000-0003-2240-528X}{\usebox{\ORCIDicon}}} \email{gp.brito@unesp.br}
\affiliation{Departamento de F\'isica, Universidade Estadual Paulista (Unesp), Campus Guaratinguet\'a, Av. Dr. Ariberto Pereira da Cunha, 333, Guaratinguet\'a, SP, Brasil.}
\author{Antonio D. Pereira\,\href{https://orcid.org/0000-0002-6952-2961}{\usebox{\ORCIDicon}}} \email{adpjunior@id.uff.br}
\affiliation{Instituto de F\'isica, Universidade Federal Fluminense, Campus da Praia Vermelha, Av. Litor\^anea s/n, 24210-346, Niter\'oi, RJ, Brazil}

\begin{abstract}
The Refined Gribov-Zwanziger (RGZ) action in the Landau gauge provides a local and renormalizable framework to account for the existence of infinitesimal Gribov copies in the path integral together with other relevant infrared effects such as the formation of condensates. The properties of the tree-level gluon propagator obtained in this setup has been thoroughly investigated over the past decade. It accommodates important properties seen in lattice simulations such as a finite value at vanishing momentum and positivity violation. Yet a comprehensive study about the stability of such properties against quantum corrections was lacking. In this work, we compute the gluon propagator in the RGZ scenario at one-loop order and implement an appropriate renormalization scheme in order to compare our findings with lattice data. Remarkably, the qualitative properties of the tree-level gluon propagator are preserved. In particular, the fits with lattice data show evidence for positivity violation and the existence of complex poles for SU(2) and SU(3) gauge groups. We comment on the results for the ghost propagator as well. 
\end{abstract}

\maketitle

\section{Introduction}

Understanding the confining mechanism in Yang-Mills (YM) theories and quantum chromodynamics (QCD) remains a challenging open problem in theoretical Physics \cite{Greensite:2011zz,Brambilla:2014jmp}. Yet, over the past three decades, we have improved substantially our knowledge about the low-energy behavior of correlation functions in those theories, see, e.g. \cite{Roberts:1994dr,Alkofer:2000wg,Fischer:2006ub,Binosi:2009qm,Maas:2011se,Vandersickel:2012tz,Huber:2018ned,Dupuis:2020fhh,Pelaez:2021tpq}. Since they are the building blocks of the would be observables, one hopes to identify signatures of confinement in those elementary correlation functions. However, YM theories become strongly coupled in the infrared and leaving the perturbative paradigm is mandatory rendering the computation of propagators and vertices very challenging. In fact, much of the progress achieved in the community of correlation functions in Yang-Mills theories and QCD is due to the constructive cross-fertilization between different non-perturbative methods that have different strengths and systematics. In particular, the use of functional methods as, e.g., Dyson-Schwinger and functional renormalization group equations to compute gauge-fixed correlation functions has greatly benefited from gauge-fixed lattice simulations since those can be used as a benchmark to point out the quality of the underlying truncations in the continuum methods. Conversely, functional methods can access regimes that are technically very troublesome in the lattice as in the presence of a chemical potential. Combining the strengths of each approach seems to be a fruitful path to be taken. 

Besides the intrinsic difficulties in non-perturbative computations another source of problems arise in the low-energy regime of non-Abelian gauge theories. The gauge-fixing procedure, which is a necessary technicality in computations performed in the continuum, is implemented via the so-called Faddeev-Popov (FP) method \cite{Faddeev:1967fc}. Such method is very successfully used in perturbation theory and, in fact, has remarkable properties such as the emergence of the global BRST symmetry in the gauge-fixed action \cite{Becchi:1974xu,Becchi:1974md,Becchi:1975nq,Tyutin:1975qk,Baulieu:1981sb}. Nonetheless, the working assumptions of the FP method cease to be valid in the non-perturbative regime. Hence there is no \textit{a priori} reason to rely on the method for sufficiently large values of the coupling constant. The breakdown of the FP procedure is related to the existence of field configurations that satisfy the same gauge condition and are connected by a gauge transformation. This is the so-called Gribov problem \cite{Gribov:1977wm,Vandersickel:2012tz,Sobreiro:2005ec}. Dealing with such Gribov copies corresponds to an improvement of the FP method in the infrared that might play a pivotal role to the correct description of correlation functions in non-Abelian gauge theories.

The Refined Gribov-Zwanziger (RGZ) action in the Landau gauge provides a local and renormalizable framework to incorporate the elimination of infinitesimal Gribov copies as well as the formation of dimension-two condensates in the path integral of Yang-Mills theories \cite{Dudal:2007cw,Dudal:2008sp}. In practice, the RGZ action introduces modifications to the gauge-fixed Yang-Mills action that affect the tree-level gluon propagator and new propagators as well as vertices involving auxiliary fields that are necessary for localization purposes. The RGZ gluon propagator at tree-level can be fitted to gauge-fixed lattice simulations and remarkably yields a very good description of the lattice data for sufficiently low momentum \cite{Dudal:2010tf,Cucchieri:2011ig,Dudal:2018cli}. In essence, the tree-level gluon propagator in the RGZ scenario carries massive parameters that, in principle, are not free but fixed by their own gap equations and carry non-pertubative signatures. Yet solving such gap equations self-consistently is a challenging task \cite{Dudal:2005na,Dudal:2011gd,Dudal:2019ing} and requires a good knowledge of radiative corrections to the RGZ theory.  

Over the past decade, it became clear that the so-called Curci-Ferrari model, which is a massive extension of the Yang-Mills action in the Landau gauge, provides an effective model that gives rise to perturbative correlation functions that compare very well with gauge-fixed lattice simulations \cite{Tissier:2010ts,Tissier:2011ey,Pelaez:2013cpa,Reinosa:2017qtf,Barrios:2024ixj,Gracey:2019xom,Barrios:2020ubx}. At one-loop order already, the gluon propagator compares very well with lattice data as well as other correlation functions such as the ghost propagator, ghost-gluon and pure gluonic vertices \cite{Tissier:2010ts,Pelaez:2013cpa,Barrios:2024ixj}. The tree-level gluon propagator in the RGZ framework can be written as a sum of two Yukawa propagators, i.e., 
\begin{equation}
\mathcal{D}_{AA}(p) = \frac{R_+}{p^2 + \mu^2_{+}}+\frac{R_{-}}{p^2 + \mu^2_{-}}\,,
\label{intro1}
\end{equation}
where $\mathcal{D}_{AA}(p)$ is the gluon propagator form factor and $(\mu^2_{\pm},R_{\pm})$ are functions of the mass parameters that are introduced in the RGZ action. Such a structure suggests that a perturbative treatment of the RGZ theory, in analogy to the Curci-Ferrari model, could provide a good description of the correlation functions of elementary fields in the low momentum regime. Nevertheless, unlike the Curci-Ferrari model, the RGZ action has an intricate set of Feynman rules with mixed propagators and new vertices. This engenders the proliferation of extra diagrams and even one-loop calculations in this setup become intricate. For this reason, much of the progress in computing correlation functions in the RGZ scenario has been very slow. In particular, one-loop corrections to correlation functions in the RGZ theory were computed in \cite{Gracey:2012wf,Mintz:2017qri,Barrios:2024idr} and in \cite{deBrito:2023qfs} in the presence of scalar fields. The aim of this paper is to report on the results for the gluon and ghost propagators at one-loop order within the RGZ framework and its comparison with lattice simulations.

The paper is organized as follows: In Sect.~\ref{Sect:OverRGZ}, we present a short introduction to the RGZ framework in order to set up our conventions. Sect.~\ref{Sect:GluonPropRGZ} presents our results for the one-loop gluon propagator in the RGZ scenario both for SU($2$) and SU($3$) gauge groups. We also fit our findings with available lattice data. Sect.~\ref{Sect:GhostPropRGZ} is devoted to the ghost propagator at one-loop order and we collect our conclusions and provide an outlook in Sect.~\ref{Sect:Outlook}.

\section{A very brief overview of the RGZ framework \label{Sect:OverRGZ}}

Pure YM theories in four Euclidean dimensions and gauge group SU($N$) are described by the action\footnote{We employ the short-hand notation $\int_ x \equiv \int {\rm d}^4x$.}
\begin{equation}
S_{\rm YM} = \frac{1}{4}\int_x F^{a}_{\mu\nu}F^{a}_{\mu\nu}\,,
\label{Eq:OverRGZ.1}
\end{equation}
with the latin indices running as $a = 1,\ldots , N^2-1$. The field strength $F^{a}_{\mu\nu}$ is defined as $F^{a}_{\mu\nu} = \partial_\mu A^a_\nu - \partial_\nu A^a_\mu + gf^{abc} A^{b}_{\mu} A^c_{\nu}$ with $f^{abc}$ being the totally antisymmetric structure constants of SU($N$) and $g$ the dimensionless coupling constant of YM theories. The covariant derivative in the adjoint representation of the gauge group is defined as $D^{ab}_\mu \equiv \delta^{ab} \partial_\mu - gf^{abc}A^{c}_{\mu}$. 

The Landau gauge condition can be implemented by means of the FP procedure which amounts to supplement the action \eqref{Eq:OverRGZ.1} with the following terms,
\begin{equation}
S_{\rm FP} = \int_x \Big(b^a \partial_\mu A^a_\mu + \bar{c}^a \partial_\mu D^{ab}_\mu c^b\Big)\,.
\label{Eq:OverRGZ.2}
\end{equation}
The fields $(b,\bar{c},c)$ are, respectively, the Lautrup-Nakanishi field and the FP anti-ghost and ghost fields. Hence, the gauge-fixed YM theory in the Landau gauge $\partial_\mu A^a_\mu = 0$ is described by the action
\begin{equation}
S_{L} = S_{\rm YM} + S_{\rm FP}\,.
\label{Eq:OverRGZ.3}
\end{equation}
The FP partition function $\EuScript{Z}_{\rm FP}$ is defined as
\begin{equation}
\EuScript{Z}_{\rm FP} = \int [\EuScript{D}\mu]_{\rm FP}\,{\rm e}^{-S_{L}}\,,
\label{Eq:OverRGZ.4}
\end{equation}
with
\begin{equation}
[\EuScript{D}\mu]_{\rm FP} = [\EuScript{D}A][\EuScript{D}b][\EuScript{D}\bar{c}][\EuScript{D}c]\,.
\label{Eq:OverRGZ.5}
\end{equation}
It is well-known that the Landau gauge condition does not remove completely the gauge redundancy. Consider two gauge fields $A^a_\mu$ and $\tilde{A}^a_\mu$ that satisfy the Landau gauge condition and are connected by an infinitesimal gauge transformation, i.e.,
\begin{equation}
\tilde{A}^a_\mu = A^a_\mu - D^{ab}_\mu \theta^b\,,
\label{Eq:OverRGZ.6}
\end{equation}
with $\theta^a$ being an infinitesimal parameter of the gauge transformation. Taking the divergence of \eqref{Eq:OverRGZ.6} and imposing the Landau gauge condition yields
\begin{equation}
\partial_\mu \tilde{A}^a_\mu = \partial_\mu A^a_\mu - \partial_\mu D^{ab}_\mu\,\theta^b\,\quad \Rightarrow \quad - \partial_\mu D^{ab}_\mu\,\theta^b = 0\,.
\label{Eq:OverRGZ.7}
\end{equation}
This means that if the FP operator $-\partial_\mu D^{ab}_\mu$ has (normalizable) zero modes, then different gauge-field configurations that are related by infinitesimal gauge transformations can satisfy the Landau gauge condition. It turns out that those zero modes exist as pointed out in \cite{Gribov:1977wm}, see also \cite{Guimaraes:2011sf}. These residual spurious gauge configurations are known as Gribov copies. In fact, the existence of Gribov copies, i.e., the Gribov problem, is not restricted to configurations that are infinitesimally related, but it occurs also in the case of finite gauge transformed fields \cite{vanBaal:1991zw}. Moreover, this is not a pathology of the Landau gauge, but a geometrical obstruction that is ubiquitous in non-Abelian gauge theories \cite{Singer:1978dk}.

The presence of Gribov copies suggests that the FP method should be improved in order to remove those residual configurations. In the case of the Landau gauge, such an improvement was thoroughly investigated, see, \cite{Zwanziger:1989mf,Vandersickel:2012tz,Zwanziger:1988jt}, leading to the so-called Gribov-Zwanziger (GZ) action, which eliminates infinitesimal Gribov copies and is given by\footnote{The GZ action extends to all orders in perturbation theory the analysis performed by Gribov in \cite{Gribov:1977wm} at leading order. The strategy followed by Zwanziger in \cite{Zwanziger:1989mf} is not the same as the one carried out by Gribov in \cite{Gribov:1977wm}. Nonetheless, they lead to equivalent results in the Landau gauge as proved in \cite{Capri:2012wx}.}
\begin{equation}
\EuScript{Z}_{\rm GZ} = \int [\EuScript{D}\mu]_{\rm GZ}\,{\rm e}^{-S_{\rm GZ}+4\gamma^4(N^2-1)V}\,,
\label{Eq:OverRGZ.8}
\end{equation}
with
\begin{equation}
S_{\rm GZ} = S_{\rm YM} + S_{\rm FP} + S_H\,,
\label{Eq:OverRGZ.9}
\end{equation}
and
\begin{eqnarray}
S_H &=& \int_{x}\Big(\bar{\varphi}^{ab}_\mu\,\EuScript{M}^{ac}(A)\,\varphi^{cb}_\mu-\bar{\omega}^{ab}_\mu\,\EuScript{M}^{ac}(A)\,\omega^{cb}_\mu\Big)\nonumber\\
&+&ig\gamma^2 \int_{x} f^{abc}A^a_\mu (\varphi^{bc}_\mu + \bar{\varphi}^{bc}_\mu)\,.
\label{Eq:OverRGZ.10}
\end{eqnarray}
The functional measure is written as $[\EuScript{D}\mu]_{\rm GZ} = [\EuScript{D}A][\EuScript{D}b][\EuScript{D}\bar{c}][\EuScript{D}c][\EuScript{D}{\bar \varphi}][\EuScript{D}\varphi][\EuScript{D}\bar{\omega}][\EuScript{D}\omega]$. The parameter $\gamma^2$ is the so-called Gribov parameter. It is not free but fixed by a gap equation,
\begin{equation}
\frac{\partial \EuScript{E}_v}{\partial \gamma^2}\Bigg|_{\gamma^2 \neq 0} = 0\,,
\label{Eq:OverRGZ.11}
\end{equation}
with $\EuScript{E}_v$ being the vacuum energy defined as ${\rm e}^{-V\EuScript{E_v}} = \EuScript{Z}_{\rm GZ}$ with $V$ denoting the spacetime volume. The fields $(\bar{\varphi},\varphi)^{ab}_\mu$ and $(\bar{\omega},\omega)^{ab}_\mu$ are, respectively, bosonic and Grassmannian auxiliary fields that can be integrated out rendering a non-local action. 
In the following, we replace the Gribov parameter by a massive parameter $\lambda$ such that
\begin{equation}
	ig\gamma^2 \!\! \int_{x}\!\! f^{abc}A^a_\mu (\varphi^{bc}_\mu + \bar{\varphi}^{bc}_\mu)
	=
	\frac{i \,\lambda^2}{\sqrt{2N}} \! \int_{x}\!\! f^{abc}A^a_\mu (\varphi^{bc}_\mu + \bar{\varphi}^{bc}_\mu) \,.
\end{equation}
Expressing the GZ action in terms of $\lambda$ instead of $\gamma$ allows us to establish a correspondence between the loop expansion with the expansion in powers of $g^2$. 

Effectively, the GZ action implements the restriction of the path integral to the so-called Gribov region $\Omega$ in field space. This region is defined by
\begin{equation}
\Omega = \left\{A^a_\mu\,,\,\,\partial_\mu A^a_\mu = 0\,|\, -\partial_\mu D^{ab}_\mu > 0\right\}\,.
\label{Eq:OverRGZ.12}
\end{equation}
Within the Gribov region\footnote{The Gribov region displays remarkable properties: It is bounded in every direction, it is convex, and every gauge orbit crosses it at least once \cite{DellAntonio:1991mms}.} $\Omega$, the FP operator $-\partial_\mu D^{ab}_\mu$ is positive and therefore does not develop zero modes. Hence, from Eq.\eqref{Eq:OverRGZ.7}, it is clear that the Gribov region is free of infinitesimal Gribov copies. Yet this region is not entirely free of Gribov copies due to the presence of those generated by finite gauge transformations as verified in \cite{vanBaal:1991zw}. A region completely free of Gribov copies, the so-called Fundamental Modular Region exists, but a practical implementation of the restriction of the path integral to it remains unknown. The GZ action implements the restriction of the path integral to $\Omega$ in a local and renormalizable fashion \cite{Zwanziger:1989mf}.

One of the remarkable features of the GZ action is that the gluon propagator at the tree-level is profoundly modified at low momentum scales. In particular,
\begin{equation}
\langle A^a_\mu (p) A^b_\nu (-p)\rangle^{\rm GZ}_0 = \frac{p^2}{p^4+\lambda^4}\,\delta^{ab}\EuScript{P}_{\mu\nu}(p)\,,
\label{Eq:OverRGZ.13}
\end{equation}
where $\EuScript{P}_{\mu\nu}(p)$ stands for the transverse projector. The gluon propagator vanishes at vanishing momentum (this fact is an exact property of the gluon propagator in the GZ framework and not just a tree-level artifact). This fact ensures that the gluon propagator violates reflection positivity and therefore one is not able to assign a physical interpretation to the gluon as a particle in the physical spectrum. This is often interpreted as evidence for confinement. Notice that, from \eqref{Eq:OverRGZ.13}, the gluon propagator displays purely (complex conjugate) imaginary poles. At leading order in a perturbative treatment of the ghost propagator with the Feynman rules generated by the GZ action, the ghost propagator is enhanced in the deep infrared and thus the GZ action offers a scaling solution \cite{Kugo:1979gm,vonSmekal:1997ohs,vonSmekal:1997ern} for the propagator of the elementary fields of gauge-fixed Yang-Mills theories, see also \cite{Zwanziger:2001kw,Gracey:2006dr,Huber:2009tx}.

In \cite{Dudal:2007cw,Dudal:2008sp}, it was pointed out that further non-perturbative effects can be accounted for in the GZ paradigm. The inclusion of dimension-two condensates arising both from the gluonic sector as well as the auxiliary localizing fields gave birth to the so-called Refined GZ (RGZ) action, namely, 
\begin{equation}
S_{\rm RGZ} = S_{\rm GZ} + S_{\rm cond}
\label{Eq:OverRGZ.14}
\end{equation} 
with
\begin{equation}
S_{\rm cond} = \int_{x}\left[\frac{m^2}{2}A^a_\mu A^a_\mu+M^2(\bar{\varphi}^{ab}_\mu \varphi^{ab}_\mu - \bar{\omega}^{ab}_\mu \omega^{ab}_\mu)\right]\,.
\label{Eq:OverRGZ.15}
\end{equation}
The condensates masses $m^2$ and $M^2$ are not free but determined by their own gap equations, see, e.g., \cite{Dudal:2011gd,Dudal:2019ing}. The refining condensates affect the tree-level gluon propagator as follows,
\begin{equation}\label{Eq:TreeProp1}
\langle A_\mu^a(p) A_\nu^b(-p) \rangle^{\rm RGZ}_0 = \mathcal{D}^\text{tree}_{AA}(p)\,\delta^{ab}\EuScript{P}_{\mu\nu}\,,
\end{equation}
with
\begin{equation}\label{Eq:TreeProp2}
\mathcal{D}_{AA}^\text{tree}(p) = \frac{p^2 + M^2}{\left(p^2 + M^2\right) \left(p^2 + m^2\right) + \lambda^4}\,.
\end{equation}
The tree-level gluon propagator \eqref{Eq:TreeProp2} does not vanish at zero momentum and the ghost propagator, at one-loop order, is not enhanced in the infrared as in the case of the GZ theory. Such a behavior is observed in lattice simulations with reasonably large lattices and gives rise to the so-called massive or decoupling solution \cite{Cucchieri:2007rg,Cucchieri:2008fc,Bornyakov:2008yx,Bogolubsky:2009dc}. Such a solution has been found in different non-perturbative approaches to the computation of infrared propagators in Yang-Mills theories such as Dyson-Schwinger equations and the Functional Renormalization Group, see \cite{Aguilar:2008xm,Fischer:2008uz,Alkofer:2008jy}. Thus the RGZ scenario corresponds to an effective model which takes into account first-principles issues such as the Gribov problem and infrared instabilities such as the formation of condensates already at the starting point action. In principle, the model does not introduce new parameters with respect to those already present in Yang-Mills theories since the mass-parameters are fixed by gap equations. Yet this is a very difficult task and one can fit those parameters with lattice data for some correlation function, such as the gluon propagator. So far, much of the progress in this context is restricted to the use of the tree-level gluon propagator in the Landau gauge and it is already able to provide a qualitative description of the lattice data, see \cite{Dudal:2010tf,Cucchieri:2011ig,Dudal:2018cli}. Yet it remains open to verify whether such successful results at the tree-level remain valid upon the inclusion of loop corrections to the correlation functions. This is the topic of the next section. 

As a final remark, the RGZ theory as just presented suffers from a major drawback: The RGZ action breaks BRST symmetry explicitly but softly. This issue has been deeply investigated for many years, see \cite{Maggiore:1993wq,Baulieu:2008fy,Dudal:2009xh,Sorella:2009vt,Sorella:2010it,Capri:2010hb,Lavrov:2011wb,Serreau:2012cg,Serreau:2013ila,Dudal:2012sb,Pereira:2013aza,Pereira:2014apa,Lavrov:2013boa,Capri:2014bsa,Cucchieri:2014via,Moshin:2014xka,Schaden:2014bea,Schaden:2015uua}. In \cite{Capri:2015ixa}, a BRST-invariant formulation of the RGZ action was proposed and this has cleared the way to establish the consistency of the results obtained in the BRST-broken version in the Landau gauge. In \cite{Capri:2015nzw,Capri:2016aqq,Capri:2016gut,Capri:2017bfd,Capri:2018ijg}, such BRST-invariant formulation was established as a local and renormalizable framework to extend the RGZ scenario beyond the Landau gauge. A consequence of such a construction is that the gluon-propagator in the Landau gauge is equivalent to the two-point function of a BRST-invariant composite field $A^h_\mu$, i.e., 
\begin{equation}
\langle A^{h,a}_\mu (x)A^{h,b}_\nu (y)\rangle = \langle A^{a}_\mu (x)A^{b}_\nu (y)\rangle_{\mathrm{Landau}}\,.
\end{equation}
Hence, the knowledge of the gluon propagator in the Landau gauge also provides insights on a gauge-invariant correlation function in this BRST-invariant construction.

\section{The Gluon Propagator \label{Sect:GluonPropRGZ}}
\begin{figure*}[t!]
	\includegraphics[width=.95\linewidth]{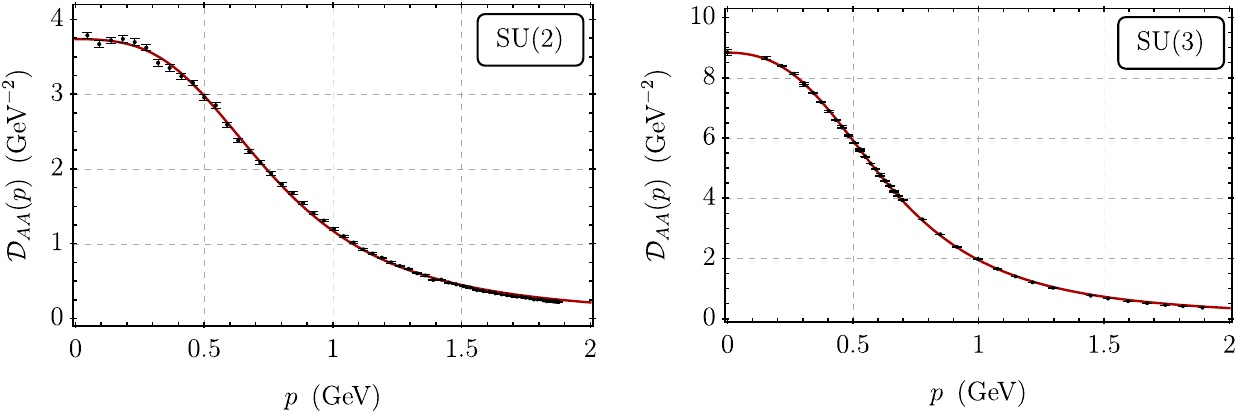}
	\caption{One-loop results for the gluon propagator in the RGZ model (full line), both for SU(2) (left-panel) and SU(3) (right-panel) gauge groups. The dots correspond to the lattice data reported in \cite{Cucchieri:2007rg,Duarte:2016iko}.
	In this plot, we fix the RGZ parameters according to the best-fitting values reported in Tab. \ref{tab:bestfitvalues}.}
	\label{fig:PropGluon}
\end{figure*}
\begin{figure*}[t!]
	\includegraphics[width=.95\linewidth]{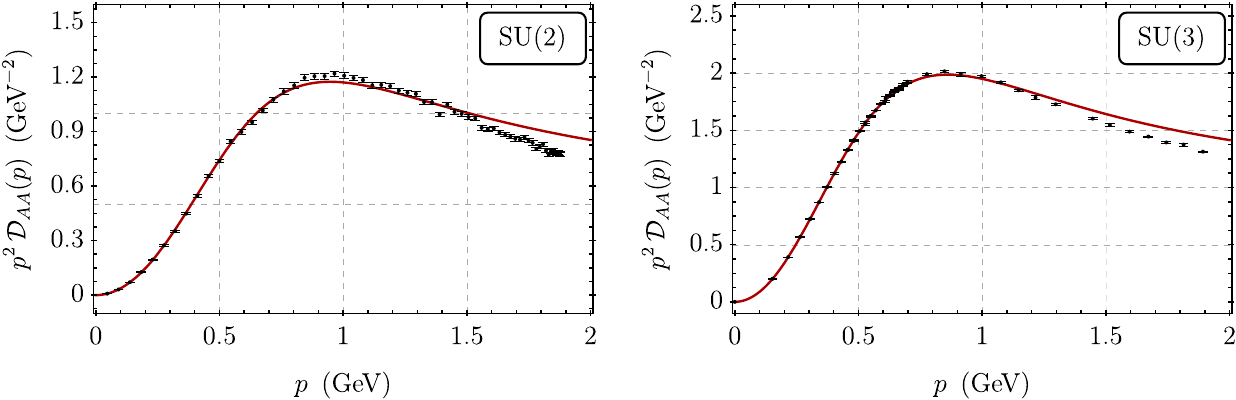}
	\caption{RGZ gluon propagator multiplied by $p^2$, both for SU(2) (left-panel) and SU(3) (right-panel) gauge groups. The dots correspond to the lattice data reported in \cite{Cucchieri:2007rg,Duarte:2016iko}.
	We fix the RGZ parameters according to the best-fitting values reported in Tab. \ref{tab:bestfitvalues}.}
	\label{fig:p2PropGluon}
\end{figure*}
\begin{figure}[t!]
	\includegraphics[width=.95\linewidth]{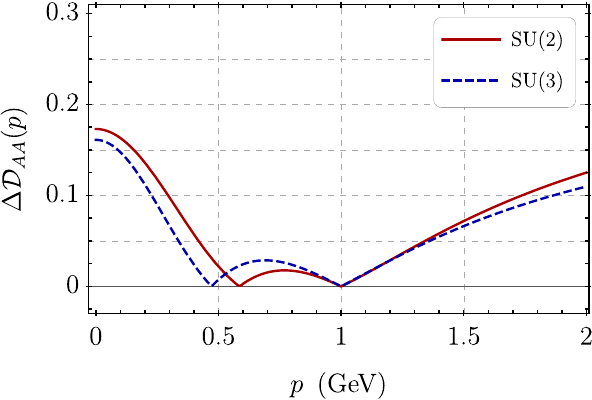}
	\caption{
	We show the relative deviation (c.f.~Eq.~\eqref{eq:DeltaDAA}) between the one-loop and tree-level results for the gluon propagator in the RGZ for gauge groups SU($2$) and SU($3$).
	}
	\label{fig:DeltaPropGluon}
\end{figure}
\begin{figure}[t!]
	\includegraphics[width=.95\linewidth]{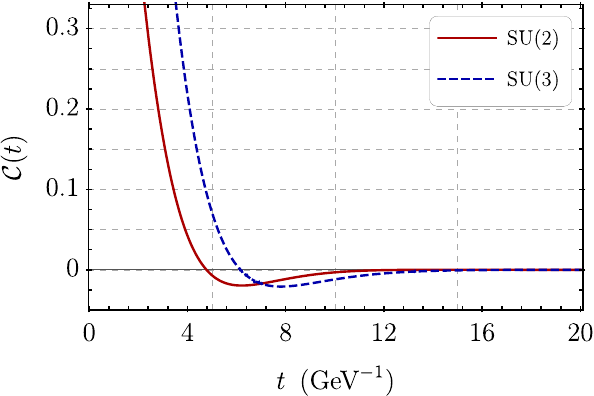}
	\caption{
	We show the temporal correlator (c.f.~Eq.~\eqref{eq:SchwingerFunc}) obtained by Fourier transforming the RGZ gluon propagator at one-loop. As one can see, the Schwinger function shows indications of positivity violation.
	}
	\label{fig:SchwingerFunc}
\end{figure}

In this section, we present the first results for the gluon propagator beyond tree level in the RGZ model.  Our focus is the one-loop correction to the connected two-point correlation function $\langle A_{\mu}^a(p) A_{\nu}^b(-p)\rangle$.

Despite being ``just'' a one-loop calculation, the derivation of the results presented in this work exhibits several complications that are not present in the standard perturbative calculation of  $\langle A_{\mu}^a(p) A_{\nu}^b(-p)\rangle$ from the YM action. First, in the local formulation of the RGZ action there are mixing terms involving the products of gauge field $A_{\mu}^a$ with the bosonic auxiliary fields $\varphi$ and $\bar{\varphi}$. As a consequence, the gluon propagator does not correspond to the simple inversion of the one-particle irreducible (1PI) correlator $\Gamma_{AA}^{(2)}(p)$. Instead, it is one of the components of the propagator matrix obtained by the inversion of the matrix-valued 1PI two-point function $\Gamma^{(2)}(p)$. Writing $\Gamma^{(2)}(p) = S^{(2)} + \Delta\Gamma^{(2)}(p) + \mathcal{O}(\text{2-loops})$, where $S^{(2)}$ and $\Delta\Gamma^{(2)}(p)$ denote the tree-level and one-loop contributions to $\Gamma^{(2)}$, we can write (schematically) 
\begin{equation}
	\begin{aligned}
		\langle AA \rangle &= \langle AA \rangle_0 + \langle AA \rangle_0  \Delta\Gamma^{(2)}_{AA} \langle AA \rangle_0 \\
		&+\langle A \varphi \rangle_0 \Delta\Gamma^{(2)}_{\varphi A} \langle A A\rangle_0
		+\langle A \bar\varphi \rangle_0 \Delta\Gamma^{(2)}_{\bar\varphi A} \langle A A\rangle_0 \\
		&+\langle A A \rangle_0 \Delta\Gamma^{(2)}_{A\varphi} \langle \varphi A\rangle_0
		+\langle A A \rangle_0 \Delta\Gamma^{(2)}_{A\bar\varphi} \langle \bar\varphi A\rangle_0\\
		&+\langle A \varphi \rangle_0 \Delta\Gamma^{(2)}_{\varphi\varphi} \langle \varphi A\rangle_0
		+\langle A \varphi \rangle_0 \Delta\Gamma^{(2)}_{\varphi\bar\varphi} \langle \bar\varphi A\rangle_0 \\
		&+\langle A \bar\varphi \rangle_0 \Delta\Gamma^{(2)}_{\bar\varphi\varphi} \langle \varphi A\rangle_0
		+\langle A \bar\varphi \rangle_0 \Delta\Gamma^{(2)}_{\bar\varphi\bar\varphi} \langle \bar\varphi A\rangle_0 \\
		&+\mathcal{O}(\text{2-loops}) \,,
	\end{aligned}
\end{equation}
where $\langle \cdots \rangle_0$ represents tree-level propagators. Note that we omit possible contributions involving the Lautrup-Nakanishi field as they vanish at one-loop order in the Landau gauge. Thus, as one can see in the expression above, the one-loop correction to the gluon propagator requires the calculation of extra 1PI two-point functions that are not present in the standard YM theory.

The second complication is the appearance of extra diagrams due to the additional vertices involving the gauge field and the RGZ auxiliary fields. While the one-loop correction to the gluon propagator in the standard (pure) YM setting involves only three diagrams, the corresponding calculation in the RGZ setting involves 20 diagrams. To overcome the complexity of such calculations we have used a self-written \textit{Mathematica} \cite{Mathematica} code based on the packages \textit{xAct} \cite{xAct1,xAct2,xAct3}, \textit{FeynCalc} \cite{FeynCalc1,FeynCalc2,FeynCalc3}, \textit{DoFun} \cite{DoFun1,DoFun2} and \textit{FormTracer} \cite{FormTracer}.

It is worth noting that one can reduce the number of diagrams by working in the non-local version of the RGZ action, i.e., by integrating out the auxiliary fields. However, the complexity of the calculation re-appears in the form of larger and non-local gluon self-interaction vertices. Although in this work we focus in the local version of the RGZ, we have explicitly checked that the non-local RGZ action leads to the same results for the regularized gluon propagator at one-loop level. However, the renormalization of the non-local version has extra subtleties which we plan to discuss in a future publication.

Before presenting our main results, let us briefly discuss the renormalization scheme employed in our calculation.
In principle, the renormalization of two-point functions in the RGZ setting requires us to fix seven renormalization factors: $Z_A$, $Z_c$, $Z_\varphi$ and $Z_\omega$, corresponding to the wave function renormalization of the gluon, ghost and the RGZ auxiliary fields; $Z_{m^2}$, $Z_{M^2}$ and $Z_{\lambda^2}$, corresponding to the renormalization of the RGZ parameters $m^2$, $M^2$ and $\lambda^2$. However, the RGZ non-renormalization theorems allow us to relate the different renormalization factors \cite{Dudal:2008sp}, resulting on just two independent ones, which one can choose to be $Z_A$ and $Z_c$.  

To fix $Z_A$ and $Z_c$, we need to impose renormalization conditions.
Since our goal is to compare our analytical results with lattice data, the standard $\overline{\text{MS}}$-scheme is not suitable for our analysis. Instead, we employ the renormalization condition
\begin{subequations}
	\begin{equation}
		\langle A_{\mu}^a(-p) A_{\nu}^b(p) \rangle\big|_{p^2 = \mu^2} = 
		\langle A_{\mu}^a(-p) A_{\nu}^b(p) \rangle_0\big|_{p^2 = \mu^2} \,,
	\end{equation}
	\begin{equation}
		\langle c^a(-p) \bar{c}^b(p) \rangle\big|_{p^2 = \mu^2} = 
		\langle c^a(-p) \bar{c}^b(p) \rangle_0\big|_{p^2 = \mu^2} \,,
	\end{equation}
\end{subequations}
where $\mu$ is the renormalization scale.
Once we fix $Z_A$ and $Z_c$, we get UV-finite expressions for all one-loop two-point correlation functions. Unfortunately, the resulting analytical expressions are too involved to be reported here. We provide the relevant pieces for our analysis in an ancillary \textit{Mathematica} notebook.
Notably, the above renormalization condition allows us to reproduce the gluon and ghost propagators of the Curci-Ferrari model in the infrared-safe scheme, see, e.g., \cite{Tissier:2011ey}, by taking the limit $\lambda \to 0$ in our expressions, thus providing a consistency check of our results\footnote{The limit $\lambda \to 0$ leads to a perfect cancellation between the contribution arising from the integration of the $(\bar{\varphi},\varphi)$ fields against the integration of the $(\bar{\omega},\omega)$ fields. Hence, there is no need to take the limit $M^2\to 0$ in order to recover the Curci-Ferrari model action.}.

In the following, we present our main results for the gluon propagator at one-loop order. To facilitate our analysis, we recall that all the non-trivial information is encoded in a single scalar function $\mathcal{D}_{AA}(p)$, as the Landau gauge allow us to write
\begin{equation}
	\langle A_\mu^a(p) A_\nu^b(-p) \rangle^{\rm RGZ} = \mathcal{D}_{AA}(p)\,\delta^{ab}\EuScript{P}_{\mu\nu} \,,
\end{equation}
to all orders in perturbation theory. Besides the usual dependence on the non-Abelian gauge coupling $g$, the one-loop expression for  $\mathcal{D}_{AA}(p)$ also depends on the mass parameters $\lambda^2$, $m^2$ and $M^2$. In principle, one could fix the RGZ parameters $(\lambda^2,m^2,M^2)$ by solving their corresponding gap equation. This procedure, despite being necessary for the self-consistency of the theory, requires us to deal with local composite operators, which goes beyond the scope of the present paper. Here, we adopt a more exploratory route by treating $(\lambda^2,m^2,M^2)$ as free parameters and fixing their values from lattice data. In particular, we use two sets of lattice data for the gluon propagator, from simulations done with SU(2) \cite{Cucchieri:2007rg} and SU(3) \cite{Duarte:2016iko} gauge groups. 

In the SU(2) case, we use the data set obtained from simulations with lattice volume $V =128^4$ and $\beta=2.2$. In our analysis, we disregard the running\footnote{We fix the renormalization scale $\mu= 1\, \text{GeV}$.} of $g$ and we fixed it according to the relation $\beta=2 N/g^2$ (with $N=2$). In the SU(3) case, we use the data set obtained from simulations with lattice volume $V =80^4$ and $\beta=6.0$. Once again, we disregard the running of $g$, fixing it according to the relation $\beta=2 N/g^2$ (with $N=3$). We also include a multiplicative parameter $\mathcal{A}$ to account for the normalization of the lattice data. Therefore, the fitting procedure is actually done with $\tilde{\mathcal{D}}_{AA}(p;\mathcal{A},\lambda^2,m^2,M^2) = \mathcal{A}\,\mathcal{D}_{AA}(p;\lambda^2,m^2,M^2)$. For simplicity of notation, we drop the tilde in the rest of this paper.

In Fig.~\ref{fig:PropGluon}, we plot our results for the fitting of the gluon propagator based on SU(2) (left-panel) and SU(3) (right-panel). We report the corresponding best-fit values in Tab. \ref{tab:bestfitvalues}.

\begin{table}[t]
	\begin{tabular}{|c|c|c|c|c|}\hline\hline
								  & $\,\lambda^2\,(\text{GeV}^2)$ & $\,m^2\,(\text{GeV}^2)$ & $\,M^2\,(\text{GeV}^2)$  & $\quad \mathcal{A} \quad $ \\\hline\hline
		SU(2), tree & $5.65$ & $-5.35$ & $5.86$ & $0.36$ \\\hline
		SU(2), one-loop & $ 2.07$ & $-1.63$ & $2.39$ & $0.74$\\\hline\hline
		SU(3), tree & $3.54$ & $-3.23$ & $3.79$  & $0.77$ \\\hline
		SU(3), one-loop & $1.90$ & $-1.47$ & $2.28$ & $1.23$\\\hline\hline	
	\end{tabular}
	\caption{Best-fit values based on comparison between the gluon propagator (tree-level and one-loop) in RGZ model and lattice results.}
	\label{tab:bestfitvalues}
\end{table}

First, we observe that despite the highly non-trivial form of the one-loop correction to the gluon propagator in the RGZ setting, we can still choose the RGZ parameters $\lambda^2$, $m^2$ and $M^2$ in such a way that we can accommodate the lattice data in the infrared region quite accurately. The propagators at one-loop order, both in SU(2) and SU(3), attain a finite value at vanishing momentum and they read

\begin{subequations}
\begin{equation}
\mathcal{D}_{AA} (0)\big|_\text{SU(2)}  \approx 3.74\,\mathrm{GeV}^{-2}\,,\\
\end{equation}
\begin{equation}
\mathcal{D}_{AA} (0)\big|_\text{SU(3)}  \approx 8.83\,\mathrm{GeV}^{-2}\,.
\end{equation}
\end{subequations}
In Fig.~\ref{fig:p2PropGluon}, we show the corresponding plots for $p^2 \mathcal{D}_{AA}(p)$, both for SU(2) (left-panel) and SU(3) (right-panel) results. These plots allow us to better visualize the agreement between our results and the lattice data in the infrared region ($p\lesssim1 \,\text{GeV}$).
The disagreement in the region $p > 1\,\text{GeV}$ is a consequence of the fact that we are not taking into account the running of the gauge coupling to tame the large logarithm corrections. The inclusion of such renormalization group effects will be reported elsewhere.

Second, we note that the best-fit values of the RGZ parameters lie within a region where the one-loop correction to the gluon propagator is small in comparison with the tree-level expression. To verify this feature we define the ``relative deviation''
\begin{equation}\label{eq:DeltaDAA}
	\Delta_{AA}(p) = \bigg|\frac{ \mathcal{D}_{AA}(p) - \mathcal{D}^\text{tree}_{AA}(p)}{\mathcal{D}^\text{tree}_{AA}(p)}\bigg|_\text{one-loop fit}\,,
\end{equation}
where the subscript ``one-loop fit'' indicates that the RGZ parameters are fixed by their best-fit values obtained from our one-loop result.
In Fig.~\ref{fig:DeltaPropGluon}, we plot $\Delta_{AA}(p)$ for the SU(2) and SU(3) cases. As we can see, the relative deviation $\Delta_{AA}(p)$ is smaller than 20\% for all values of $p$ within the plot range, including the deep infrared regime.

The tree-level gluon propagator in the RGZ model has two appealing features, frequently associated as an indication of gluon confinement, see, e.g., \cite{Alkofer:2000wg} and \cite{Cucchieri:2011ig}. First, the gluon propagator has two complex poles. Second, the gluon propagator violates positivity. One can check that both features are also present for the one-loop gluon propagator.

To determine the pole structure of the one-loop gluon propagator we note that, to first order in $g^2$, one can rewrite the gluon propagator as
\begin{equation}
	\mathcal{D}_{AA}(p) = \frac{1}{\big( \mathcal{D}^\text{tree}_{AA}(p) \big)^{-1} - g^2 \, \Sigma_{AA}(p)} \,,
\end{equation}
where $g^2 \, \Sigma_{AA}(p)$ can be computed by matching the $g^2$-term of the above expression with the one-loop result previously computed. Restricting ourselves to corrections up to $\mathcal{O}(g^2)$, we can approximate the propagator pole as
\begin{equation}
	p^2_{\pm} \approx p^2_{0,\pm} + g^2 \, p^2_{1,\pm} \,,
\end{equation} 
where $p^2_{0,\pm}$ correspond to the poles of the tree-level gluon propagator, namely
\begin{align}
	p^2_{0,\pm} = - \frac{m^2 + M^2 \pm \sqrt{(m^2 - M^2)^2 - 4 \lambda^4} }{2} \,.
\end{align}
The one-loop contribution is given by
\begin{equation}
	p^2_{1,\pm} = - \Bigg[ \left( \frac{\partial \big( \mathcal{D}^\text{tree}_{AA}(p) \big)^{-1}}{ \partial p^2} \right)^{\,-1} \, \Sigma_{AA}(p) \Bigg]_{p^2 = p^2_{0,\pm}} \,.
\end{equation}
Based on our fitting for gluon propagator, we obtain
\begin{subequations}
	\begin{equation}
		p^2_{\pm}|_\text{SU(2)} \approx 
		-0.184 \pm 0.667 \,i \,\,\,\text{GeV}^{2}\,,
	\end{equation} 
	\begin{equation}
		p^2_{\pm}|_\text{SU(3)} \approx 
		-0.404 \pm 0.322 \, i \,\,\,\text{GeV}^{2}\,.
	\end{equation} 
\end{subequations}
It is important to emphasize that the poles are complex conjugate numbers. Unlike the case of the GZ theory where the poles are purely imaginary, the present poles have a non-vanishing real part signaling a departure from the GZ propagator. 

To identify the positivity violating nature of the gluon propagator, we look at the temporal correlator defined as, see, e.g., \cite{Alkofer:2000wg}
\begin{equation}\label{eq:SchwingerFunc}
	\mathcal{C}(t) = \frac{1}{2\pi}\int_{-\infty}^{\infty} \!\!\mathrm{d}\omega \, \mathcal{D}_{AA}(\omega) \,e^{i\omega t} \,.
\end{equation} 
Given the intricate nature of our one-loop result for the gluon propagator, one has to evaluate the above integral numerically. In Fig.~\ref{fig:SchwingerFunc}, we plot the Schwinger function obtained by numerical integration with best-fitting parameters based on SU(2) (left-panel of Fig.~\ref{fig:PropGluon}) and SU(3) (right-panel of Fig.~\ref{fig:PropGluon}) lattice simulations. In both cases, one can see that $\mathcal{C}(t)$ has become negative around $t = 5\, \text{GeV}^{-1}$. This confirms the positivity violation of the gluon propagator in the RGZ setting at one-loop order.

\section{The Ghost propagator \label{Sect:GhostPropRGZ}}
\begin{figure*}[t!]
	\includegraphics[width=.95\linewidth]{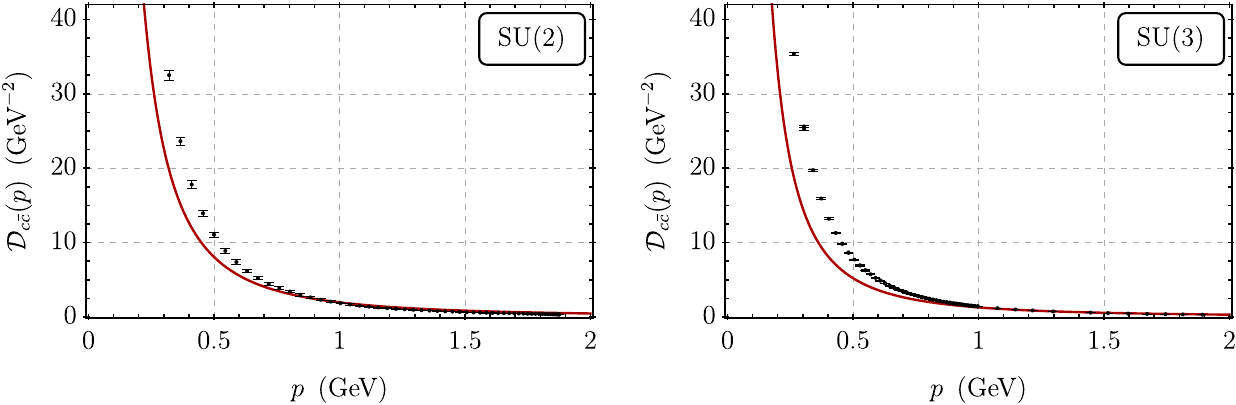}
	\caption{
	One-loop result for the ghost propagator in the RGZ setting (full line). We show the results obtained from SU(2) (left-panel) and SU(3) (right-panel). The dots correspond to the lattice data reported in \cite{Cucchieri:2007rg,Duarte:2016iko}.
	}
	\label{fig:PropGhost}
\end{figure*}
\begin{figure*}[t!]
	\includegraphics[width=.95\linewidth]{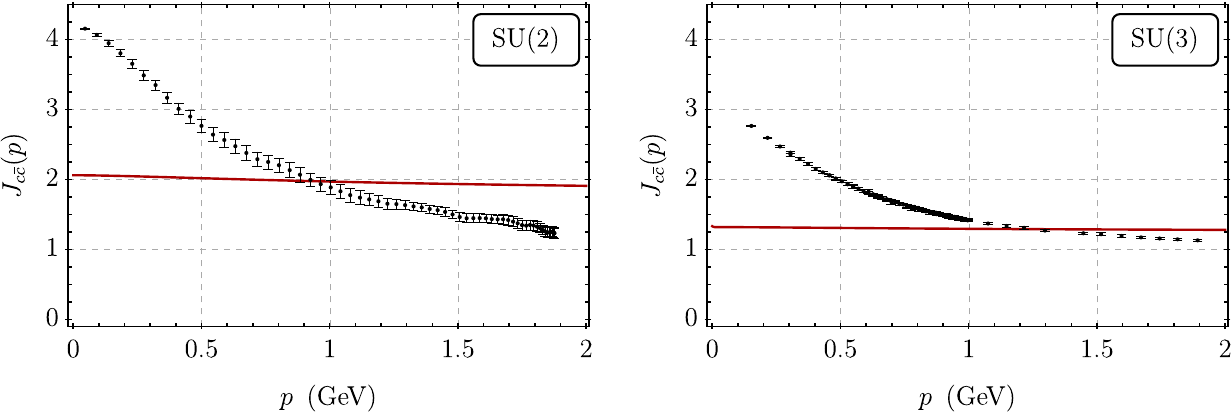}
	\caption{
	One-loop result for the ghost dressing function in the RGZ setting (full line). We show the results obtained from SU(2) (left-panel) and SU(3) (right-panel). The dots correspond to the lattice data reported in \cite{Cucchieri:2007rg,Duarte:2016iko}.
	}
	\label{fig:JpropGhost}
\end{figure*}

In this section, we revisit the ghost propagator at one loop in light of the new results for the gluon propagator. First, it is worth noticing that the one-loop correction to the ghost propagator is much simpler to compute in comparison with the gluon propagator. There is a single one-loop diagram contributing to the ghost propagator in the RGZ setting, namely, the sunset diagram with one ghost and one gluon propagator as internal lines. The difference with respect to the usual YM theory is that in the RGZ setting the gluon internal line takes the form of the tree-level propagator in Eqs. \eqref{Eq:TreeProp1} and \eqref{Eq:TreeProp2}. Following the standard notation, we express the ghost propagator as
\begin{equation}
	\langle c^a(p) \bar{c}^b(-p) \rangle = -\mathcal{D}_{c\bar{c}}(p)\, \delta^{ab} = -\frac{J_{c\bar{c}}(p)}{p^2} \,\delta^{ab} \,,
\end{equation}
where the dressing function $J_{c\bar{c}}(p)$ captures the quantum corrections. The minus sign is due to our convention for the Faddeev-Popov action, c.f.~Eq.~\eqref{Eq:OverRGZ.2}.

Once again, our goal is to compare our one-loop results with the lattice data based on simulations with SU(2) \cite{Cucchieri:2007rg} and SU(3) \cite{Duarte:2016iko} gauge groups.
Since we already fixed the RGZ parameters from the gluon propagator, we simply replace their values into the ghost propagator as input parameters. In this case, the only parameter we need to fit by comparison with the lattice data is a global constant $\mathcal{A}_\text{gh}$ introduced to account for the renormalization of the lattice data. We find the best fitting values
\begin{align}
	&\mathcal{A}_\text{gh}\big|_\text{SU(2)} = 1.97 \,,\\
	&\mathcal{A}_\text{gh}\big|_\text{SU(3)} = 1.29 \,.
\end{align}

In Fig. \ref{fig:PropGhost}, we plot our results for the ghost propagator at one-loop in comparison with the lattice data. The left panel shows the results for the SU(2) gauge group, while the right panel shows the results for the SU(3) gauge group. In Fig. \ref{fig:JpropGhost}, we plot the corresponding dressing function $J_{c\bar{c}}(p)$. In contrast with the gluon propagator, the ghost propagator in the RGZ setting does not show a good agreement with the lattice data. 

This inconsistency between the RGZ and lattice results is not a complete novelty. This was first observed in \cite{Cucchieri:2016jwg}, but using the values for the RGZ parameters obtained from the tree-level fit. Thus, in the present work, we simply confirm that such inconsistency persists if we consider the full one-loop structure of the RGZ model.

One possible direction to reconcile the RGZ and lattice results in the ghost sector is the inclusion of renormalization group effects. One expects that the running couplings play an important role in the infrared region, potentially changing the infrared behavior of the ghost propagator even at qualitative level. Unfortunately, the renormalization group analysis in the RGZ setting is quite subtle. One needs to take into account the running of the gauge coupling together with the running RGZ parameters $m^2$, $M^2$ and $\lambda^2$. Such analysis goes beyond the scope of this first work on the RGZ gluon propagator at one-loop. We plan to report on renormalization group aspects of the RGZ model in a future work. Another possibility to fix such a discrepancy in the ghost sector is to change our fitting procedure. Instead of fitting all mass parameters directly with the gluon propagator data, one could attempt to perform a combined fitting with gluon and ghost data. Unfortunately, we were not able to substantially improve the results following such a strategy in a self-consistent framework but this deserves further explorations.
	
Finally, it is conceivable that the ghost propagator at one-loop is not sufficient to capture the correct behavior of the lattice data and higher-order corrections should be taken into account \cite{Dudal:2012zx}. In particular, the ghost-gluon vertex was studied in the RGZ framework \cite{Mintz:2017qri,Barrios:2024idr} and could be added as a first improvement of the one-loop computations to the ghost sector.
	
To conclude this section, a clarification is in order: The Curci-Ferrari model, see, e.g., \cite{Tissier:2011ey} is able to provide a good description of the lattice data already at one-loop order\footnote{It should be stressed that in the works concerning the Curci-Ferrari model, the coupling constant $g$ is also taken as a free parameter to be adjusted by lattice data.}. Hence, since the RGZ theory contains the massive model as a particular case, one can conclude that an optimized fitting procedure should, at least, provide as good results as the Curci-Ferrari model (where the other massive parameters in the RGZ should be drastically suppressed). Unfortunately, we were not able to obtain such results without enforcing the vanishing of the Gribov parameter $\lambda^4$. This might be due to the very complicated structure of the one-loop expressions in the RGZ scenario. Hopefully, future investigations will enable an optimized fitting strategy to, at least, provide results that are as good as those obtained with the Curci-Ferrari model at one-loop order.


\section{Conclusions and Outlook\label{Sect:Outlook}}

Understanding the behavior of correlation functions of YM theories in the infrared corresponds to an ambitious program in order to comprehend the confining nature of non-Abelian gauge theories. The necessity to depart from the perturbative paradigm represents a severe technical challenge. Fortunately, a lot of progress has been achieved making such approach a fruitful path to be pursued and our understanding about correlation functions in YM theories has significantly improved over the past few decades.

In this work, we have computed the gluon propagator at one-loop order in the so-called RGZ framework in the Landau gauge. As pointed out, this computation is technically challenging due to the generation of several extra diagrams with respect to those that are present in standard YM theories. Such convoluted structure is rooted on the non-local nature of the modification made in the path integral in order to restrict the functional measure to the Gribov region. It can be localized at the cost of introducing several new propagators and vertices involving the localizing fields. The explicit computation was fitted to lattice data both for SU(2) and SU(3) gauge groups. We have found that the one-loop gluon propagator in the RGZ scenario in the Landau gauge fits very well the lattice data in the infrared and preserves the qualitative features of the tree-level fitting in both gauge groups. In particular, the one-loop propagator reaches a finite value at vanishing momentum, violates positivity and displays complex poles. The similarity between the tree-level and one-loop results can be viewed as a hint that within the RGZ theory, the starting point action already encodes essential non-perturbative effects that, upon the addition of radiative corrections, the results are rather stable. Clearly, this is an interpretation that requires way more consistency checks in order to be well-grounded. Our results indicate that by fitting the gluon propagator at one-loop order to the lattice data and using the fixed parameters in the expression of the ghost propagator at one-loop order is not sufficient to reproduce the lattice data. As emphasized in Sect.~\ref{Sect:GhostPropRGZ}, this is an issue that is known in the literature and there are clear paths to be taken that are already under investigation.

The results presented in this work pave the way to a self-consistent evaluation of the gluon-propagator within the RGZ framework. As explained in Sect.~\ref{Sect:OverRGZ}, the mass parameters that emerge in the RGZ setting are not free but fixed by their own gap equation. As such, the gluon propagator (and other correlation functions) could be predicted self-consistently and a good agreement between such a result with lattice date would correspond to a very important and non-trivial check for the RGZ scenario. Solving the gap equations corresponds to an important and necessary step to be incorporated in the analysis presented in this work. Similarly, our investigation was limited to fixed couplings. Introducing the appropriate running corresponds to a fundamental next-level issue to be understood. Those issues are under investigation and will be reported elsewhere.

\section*{Acknowledgments}

The authors are grateful to Marcela Pelaez, Philipe de Fabritiis, David Dudal and Bruno Mintz for helpful discussions and to Attilio Cucchieri, Tereza Mendes, Orlando Oliveira and Paulo Silva for sharing their lattice data on the gluon propagator.
GPB was supported by the research grant (29405) from VILLUM fonden during most part of this project.
ADP acknowledges CNPq under the grant PQ-2 (312211/2022-8), FAPERJ under the “Jovem Cientista do Nosso Estado” program (E26/202.800/2019 and E-26/205.924/2022) for financial support.

\bibliography{refs}

\begin{thebibliography}{95}%
\makeatletter
\providecommand \@ifxundefined [1]{%
 \@ifx{#1\undefined}
}%
\providecommand \@ifnum [1]{%
 \ifnum #1\expandafter \@firstoftwo
 \else \expandafter \@secondoftwo
 \fi
}%
\providecommand \@ifx [1]{%
 \ifx #1\expandafter \@firstoftwo
 \else \expandafter \@secondoftwo
 \fi
}%
\providecommand \natexlab [1]{#1}%
\providecommand \enquote  [1]{``#1''}%
\providecommand \bibnamefont  [1]{#1}%
\providecommand \bibfnamefont [1]{#1}%
\providecommand \citenamefont [1]{#1}%
\providecommand \href@noop [0]{\@secondoftwo}%
\providecommand \href [0]{\begingroup \@sanitize@url \@href}%
\providecommand \@href[1]{\@@startlink{#1}\@@href}%
\providecommand \@@href[1]{\endgroup#1\@@endlink}%
\providecommand \@sanitize@url [0]{\catcode `\\12\catcode `\$12\catcode
  `\&12\catcode `\#12\catcode `\^12\catcode `\_12\catcode `\%12\relax}%
\providecommand \@@startlink[1]{}%
\providecommand \@@endlink[0]{}%
\providecommand \url  [0]{\begingroup\@sanitize@url \@url }%
\providecommand \@url [1]{\endgroup\@href {#1}{\urlprefix }}%
\providecommand \urlprefix  [0]{URL }%
\providecommand \Eprint [0]{\href }%
\providecommand \doibase [0]{https://doi.org/}%
\providecommand \selectlanguage [0]{\@gobble}%
\providecommand \bibinfo  [0]{\@secondoftwo}%
\providecommand \bibfield  [0]{\@secondoftwo}%
\providecommand \translation [1]{[#1]}%
\providecommand \BibitemOpen [0]{}%
\providecommand \bibitemStop [0]{}%
\providecommand \bibitemNoStop [0]{.\EOS\space}%
\providecommand \EOS [0]{\spacefactor3000\relax}%
\providecommand \BibitemShut  [1]{\csname bibitem#1\endcsname}%
\let\auto@bib@innerbib\@empty
\bibitem [{\citenamefont {Greensite}(2011)}]{Greensite:2011zz}%
  \BibitemOpen
  \bibfield  {author} {\bibinfo {author} {\bibfnamefont {J.}~\bibnamefont
  {Greensite}},\ }\href {https://doi.org/10.1007/978-3-642-14382-3} {\emph
  {\bibinfo {title} {{An introduction to the confinement problem}}}},\ Vol.\
  \bibinfo {volume} {821}\ (\bibinfo {year} {2011})\BibitemShut {NoStop}%
\bibitem [{\citenamefont {Brambilla}\ \emph {et~al.}(2014)\citenamefont
  {Brambilla} \emph {et~al.}}]{Brambilla:2014jmp}%
  \BibitemOpen
  \bibfield  {author} {\bibinfo {author} {\bibfnamefont {N.}~\bibnamefont
  {Brambilla}} \emph {et~al.},\ }\bibfield  {title} {\bibinfo {title} {{QCD and
  Strongly Coupled Gauge Theories: Challenges and Perspectives}},\ }\href
  {https://doi.org/10.1140/epjc/s10052-014-2981-5} {\bibfield  {journal}
  {\bibinfo  {journal} {Eur. Phys. J. C}\ }\textbf {\bibinfo {volume} {74}},\
  \bibinfo {pages} {2981} (\bibinfo {year} {2014})},\ \Eprint
  {https://arxiv.org/abs/1404.3723} {arXiv:1404.3723 [hep-ph]} \BibitemShut
  {NoStop}%
\bibitem [{\citenamefont {Roberts}\ and\ \citenamefont
  {Williams}(1994)}]{Roberts:1994dr}%
  \BibitemOpen
  \bibfield  {author} {\bibinfo {author} {\bibfnamefont {C.~D.}\ \bibnamefont
  {Roberts}}\ and\ \bibinfo {author} {\bibfnamefont {A.~G.}\ \bibnamefont
  {Williams}},\ }\bibfield  {title} {\bibinfo {title} {{Dyson-Schwinger
  equations and their application to hadronic physics}},\ }\href
  {https://doi.org/10.1016/0146-6410(94)90049-3} {\bibfield  {journal}
  {\bibinfo  {journal} {Prog. Part. Nucl. Phys.}\ }\textbf {\bibinfo {volume}
  {33}},\ \bibinfo {pages} {477} (\bibinfo {year} {1994})},\ \Eprint
  {https://arxiv.org/abs/hep-ph/9403224} {arXiv:hep-ph/9403224} \BibitemShut
  {NoStop}%
\bibitem [{\citenamefont {Alkofer}\ and\ \citenamefont {von
  Smekal}(2001)}]{Alkofer:2000wg}%
  \BibitemOpen
  \bibfield  {author} {\bibinfo {author} {\bibfnamefont {R.}~\bibnamefont
  {Alkofer}}\ and\ \bibinfo {author} {\bibfnamefont {L.}~\bibnamefont {von
  Smekal}},\ }\bibfield  {title} {\bibinfo {title} {{The Infrared behavior of
  QCD Green's functions: Confinement dynamical symmetry breaking, and hadrons
  as relativistic bound states}},\ }\href
  {https://doi.org/10.1016/S0370-1573(01)00010-2} {\bibfield  {journal}
  {\bibinfo  {journal} {Phys. Rept.}\ }\textbf {\bibinfo {volume} {353}},\
  \bibinfo {pages} {281} (\bibinfo {year} {2001})},\ \Eprint
  {https://arxiv.org/abs/hep-ph/0007355} {arXiv:hep-ph/0007355 [hep-ph]}
  \BibitemShut {NoStop}%
\bibitem [{\citenamefont {Fischer}(2006)}]{Fischer:2006ub}%
  \BibitemOpen
  \bibfield  {author} {\bibinfo {author} {\bibfnamefont {C.~S.}\ \bibnamefont
  {Fischer}},\ }\bibfield  {title} {\bibinfo {title} {{Infrared properties of
  QCD from Dyson-Schwinger equations}},\ }\href
  {https://doi.org/10.1088/0954-3899/32/8/R02} {\bibfield  {journal} {\bibinfo
  {journal} {J. Phys. G}\ }\textbf {\bibinfo {volume} {32}},\ \bibinfo {pages}
  {R253} (\bibinfo {year} {2006})},\ \Eprint
  {https://arxiv.org/abs/hep-ph/0605173} {arXiv:hep-ph/0605173} \BibitemShut
  {NoStop}%
\bibitem [{\citenamefont {Binosi}\ and\ \citenamefont
  {Papavassiliou}(2009)}]{Binosi:2009qm}%
  \BibitemOpen
  \bibfield  {author} {\bibinfo {author} {\bibfnamefont {D.}~\bibnamefont
  {Binosi}}\ and\ \bibinfo {author} {\bibfnamefont {J.}~\bibnamefont
  {Papavassiliou}},\ }\bibfield  {title} {\bibinfo {title} {{Pinch Technique:
  Theory and Applications}},\ }\href
  {https://doi.org/10.1016/j.physrep.2009.05.001} {\bibfield  {journal}
  {\bibinfo  {journal} {Phys. Rept.}\ }\textbf {\bibinfo {volume} {479}},\
  \bibinfo {pages} {1} (\bibinfo {year} {2009})},\ \Eprint
  {https://arxiv.org/abs/0909.2536} {arXiv:0909.2536 [hep-ph]} \BibitemShut
  {NoStop}%
\bibitem [{\citenamefont {Maas}(2013)}]{Maas:2011se}%
  \BibitemOpen
  \bibfield  {author} {\bibinfo {author} {\bibfnamefont {A.}~\bibnamefont
  {Maas}},\ }\bibfield  {title} {\bibinfo {title} {{Describing gauge bosons at
  zero and finite temperature}},\ }\href
  {https://doi.org/10.1016/j.physrep.2012.11.002} {\bibfield  {journal}
  {\bibinfo  {journal} {Phys. Rept.}\ }\textbf {\bibinfo {volume} {524}},\
  \bibinfo {pages} {203} (\bibinfo {year} {2013})},\ \Eprint
  {https://arxiv.org/abs/1106.3942} {arXiv:1106.3942 [hep-ph]} \BibitemShut
  {NoStop}%
\bibitem [{\citenamefont {Vandersickel}\ and\ \citenamefont
  {Zwanziger}(2012)}]{Vandersickel:2012tz}%
  \BibitemOpen
  \bibfield  {author} {\bibinfo {author} {\bibfnamefont {N.}~\bibnamefont
  {Vandersickel}}\ and\ \bibinfo {author} {\bibfnamefont {D.}~\bibnamefont
  {Zwanziger}},\ }\bibfield  {title} {\bibinfo {title} {{The Gribov problem and
  QCD dynamics}},\ }\href {https://doi.org/10.1016/j.physrep.2012.07.003}
  {\bibfield  {journal} {\bibinfo  {journal} {Phys. Rept.}\ }\textbf {\bibinfo
  {volume} {520}},\ \bibinfo {pages} {175} (\bibinfo {year} {2012})},\ \Eprint
  {https://arxiv.org/abs/1202.1491} {arXiv:1202.1491 [hep-th]} \BibitemShut
  {NoStop}%
\bibitem [{\citenamefont {Huber}(2020)}]{Huber:2018ned}%
  \BibitemOpen
  \bibfield  {author} {\bibinfo {author} {\bibfnamefont {M.~Q.}\ \bibnamefont
  {Huber}},\ }\bibfield  {title} {\bibinfo {title} {{Nonperturbative properties
  of Yang\textendash{}Mills theories}},\ }\href
  {https://doi.org/10.1016/j.physrep.2020.04.004} {\bibfield  {journal}
  {\bibinfo  {journal} {Phys. Rept.}\ }\textbf {\bibinfo {volume} {879}},\
  \bibinfo {pages} {1} (\bibinfo {year} {2020})},\ \Eprint
  {https://arxiv.org/abs/1808.05227} {arXiv:1808.05227 [hep-ph]} \BibitemShut
  {NoStop}%
\bibitem [{\citenamefont {Dupuis}\ \emph {et~al.}(2020)\citenamefont {Dupuis},
  \citenamefont {Canet}, \citenamefont {Eichhorn}, \citenamefont {Metzner},
  \citenamefont {Pawlowski}, \citenamefont {Tissier},\ and\ \citenamefont
  {Wschebor}}]{Dupuis:2020fhh}%
  \BibitemOpen
  \bibfield  {author} {\bibinfo {author} {\bibfnamefont {N.}~\bibnamefont
  {Dupuis}}, \bibinfo {author} {\bibfnamefont {L.}~\bibnamefont {Canet}},
  \bibinfo {author} {\bibfnamefont {A.}~\bibnamefont {Eichhorn}}, \bibinfo
  {author} {\bibfnamefont {W.}~\bibnamefont {Metzner}}, \bibinfo {author}
  {\bibfnamefont {J.~M.}\ \bibnamefont {Pawlowski}}, \bibinfo {author}
  {\bibfnamefont {M.}~\bibnamefont {Tissier}},\ and\ \bibinfo {author}
  {\bibfnamefont {N.}~\bibnamefont {Wschebor}},\ }\bibfield  {title} {\bibinfo
  {title} {{The nonperturbative functional renormalization group and its
  applications}}\ }\href {https://doi.org/10.1016/j.physrep.2021.01.001}
  {10.1016/j.physrep.2021.01.001} (\bibinfo {year} {2020}),\ \Eprint
  {https://arxiv.org/abs/2006.04853} {arXiv:2006.04853 [cond-mat.stat-mech]}
  \BibitemShut {NoStop}%
\bibitem [{\citenamefont {Pel\'aez}\ \emph {et~al.}(2021)\citenamefont
  {Pel\'aez}, \citenamefont {Reinosa}, \citenamefont {Serreau}, \citenamefont
  {Tissier},\ and\ \citenamefont {Wschebor}}]{Pelaez:2021tpq}%
  \BibitemOpen
  \bibfield  {author} {\bibinfo {author} {\bibfnamefont {M.}~\bibnamefont
  {Pel\'aez}}, \bibinfo {author} {\bibfnamefont {U.}~\bibnamefont {Reinosa}},
  \bibinfo {author} {\bibfnamefont {J.}~\bibnamefont {Serreau}}, \bibinfo
  {author} {\bibfnamefont {M.}~\bibnamefont {Tissier}},\ and\ \bibinfo {author}
  {\bibfnamefont {N.}~\bibnamefont {Wschebor}},\ }\bibfield  {title} {\bibinfo
  {title} {{A window on infrared QCD with small expansion parameters}},\ }\href
  {https://doi.org/10.1088/1361-6633/ac36b8} {\bibfield  {journal} {\bibinfo
  {journal} {Rept. Prog. Phys.}\ }\textbf {\bibinfo {volume} {84}},\ \bibinfo
  {pages} {124202} (\bibinfo {year} {2021})},\ \Eprint
  {https://arxiv.org/abs/2106.04526} {arXiv:2106.04526 [hep-th]} \BibitemShut
  {NoStop}%
\bibitem [{\citenamefont {Faddeev}\ and\ \citenamefont
  {Popov}(1967)}]{Faddeev:1967fc}%
  \BibitemOpen
  \bibfield  {author} {\bibinfo {author} {\bibfnamefont {L.}~\bibnamefont
  {Faddeev}}\ and\ \bibinfo {author} {\bibfnamefont {V.}~\bibnamefont
  {Popov}},\ }\bibfield  {title} {\bibinfo {title} {{Feynman Diagrams for the
  Yang-Mills Field}},\ }\href {https://doi.org/10.1016/0370-2693(67)90067-6}
  {\bibfield  {journal} {\bibinfo  {journal} {Phys. Lett. B}\ }\textbf
  {\bibinfo {volume} {25}},\ \bibinfo {pages} {29} (\bibinfo {year}
  {1967})}\BibitemShut {NoStop}%
\bibitem [{\citenamefont {Becchi}\ \emph {et~al.}(1974)\citenamefont {Becchi},
  \citenamefont {Rouet},\ and\ \citenamefont {Stora}}]{Becchi:1974xu}%
  \BibitemOpen
  \bibfield  {author} {\bibinfo {author} {\bibfnamefont {C.}~\bibnamefont
  {Becchi}}, \bibinfo {author} {\bibfnamefont {A.}~\bibnamefont {Rouet}},\ and\
  \bibinfo {author} {\bibfnamefont {R.}~\bibnamefont {Stora}},\ }\bibfield
  {title} {\bibinfo {title} {{The Abelian Higgs-Kibble Model. Unitarity of the
  S Operator}},\ }\href {https://doi.org/10.1016/0370-2693(74)90058-6}
  {\bibfield  {journal} {\bibinfo  {journal} {Phys. Lett. B}\ }\textbf
  {\bibinfo {volume} {52}},\ \bibinfo {pages} {344} (\bibinfo {year}
  {1974})}\BibitemShut {NoStop}%
\bibitem [{\citenamefont {Becchi}\ \emph {et~al.}(1975)\citenamefont {Becchi},
  \citenamefont {Rouet},\ and\ \citenamefont {Stora}}]{Becchi:1974md}%
  \BibitemOpen
  \bibfield  {author} {\bibinfo {author} {\bibfnamefont {C.}~\bibnamefont
  {Becchi}}, \bibinfo {author} {\bibfnamefont {A.}~\bibnamefont {Rouet}},\ and\
  \bibinfo {author} {\bibfnamefont {R.}~\bibnamefont {Stora}},\ }\bibfield
  {title} {\bibinfo {title} {{Renormalization of the Abelian Higgs-Kibble
  Model}},\ }\href {https://doi.org/10.1007/BF01614158} {\bibfield  {journal}
  {\bibinfo  {journal} {Commun. Math. Phys.}\ }\textbf {\bibinfo {volume}
  {42}},\ \bibinfo {pages} {127} (\bibinfo {year} {1975})}\BibitemShut
  {NoStop}%
\bibitem [{\citenamefont {Becchi}\ \emph {et~al.}(1976)\citenamefont {Becchi},
  \citenamefont {Rouet},\ and\ \citenamefont {Stora}}]{Becchi:1975nq}%
  \BibitemOpen
  \bibfield  {author} {\bibinfo {author} {\bibfnamefont {C.}~\bibnamefont
  {Becchi}}, \bibinfo {author} {\bibfnamefont {A.}~\bibnamefont {Rouet}},\ and\
  \bibinfo {author} {\bibfnamefont {R.}~\bibnamefont {Stora}},\ }\bibfield
  {title} {\bibinfo {title} {{Renormalization of Gauge Theories}},\ }\href
  {https://doi.org/10.1016/0003-4916(76)90156-1} {\bibfield  {journal}
  {\bibinfo  {journal} {Annals Phys.}\ }\textbf {\bibinfo {volume} {98}},\
  \bibinfo {pages} {287} (\bibinfo {year} {1976})}\BibitemShut {NoStop}%
\bibitem [{\citenamefont {Tyutin}(1975)}]{Tyutin:1975qk}%
  \BibitemOpen
  \bibfield  {author} {\bibinfo {author} {\bibfnamefont {I.~V.}\ \bibnamefont
  {Tyutin}},\ }\bibfield  {title} {\bibinfo {title} {{Gauge Invariance in Field
  Theory and Statistical Physics in Operator Formalism}},\ }\href@noop {} {\
  (\bibinfo {year} {1975})},\ \Eprint {https://arxiv.org/abs/0812.0580}
  {arXiv:0812.0580 [hep-th]} \BibitemShut {NoStop}%
\bibitem [{\citenamefont {Baulieu}\ and\ \citenamefont
  {Thierry-Mieg}(1982)}]{Baulieu:1981sb}%
  \BibitemOpen
  \bibfield  {author} {\bibinfo {author} {\bibfnamefont {L.}~\bibnamefont
  {Baulieu}}\ and\ \bibinfo {author} {\bibfnamefont {J.}~\bibnamefont
  {Thierry-Mieg}},\ }\bibfield  {title} {\bibinfo {title} {{The Principle of
  BRS Symmetry: An Alternative Approach to Yang-Mills Theories}},\ }\href
  {https://doi.org/10.1016/0550-3213(82)90454-0} {\bibfield  {journal}
  {\bibinfo  {journal} {Nucl. Phys. B}\ }\textbf {\bibinfo {volume} {197}},\
  \bibinfo {pages} {477} (\bibinfo {year} {1982})}\BibitemShut {NoStop}%
\bibitem [{\citenamefont {Gribov}(1978)}]{Gribov:1977wm}%
  \BibitemOpen
  \bibfield  {author} {\bibinfo {author} {\bibfnamefont {V.~N.}\ \bibnamefont
  {Gribov}},\ }\bibfield  {title} {\bibinfo {title} {{Quantization of
  Nonabelian Gauge Theories}},\ }\href
  {https://doi.org/10.1016/0550-3213(78)90175-X} {\bibfield  {journal}
  {\bibinfo  {journal} {Nucl. Phys.}\ }\textbf {\bibinfo {volume} {B139}},\
  \bibinfo {pages} {1} (\bibinfo {year} {1978})},\ \bibinfo {note}
  {[1(1977)]}\BibitemShut {NoStop}%
\bibitem [{\citenamefont {Sobreiro}\ and\ \citenamefont
  {Sorella}(2005)}]{Sobreiro:2005ec}%
  \BibitemOpen
  \bibfield  {author} {\bibinfo {author} {\bibfnamefont {R.}~\bibnamefont
  {Sobreiro}}\ and\ \bibinfo {author} {\bibfnamefont {S.}~\bibnamefont
  {Sorella}},\ }\bibfield  {title} {\bibinfo {title} {{Introduction to the
  Gribov ambiguities in Euclidean Yang-Mills theories}},\ }in\ \href@noop {}
  {\emph {\bibinfo {booktitle} {{13th Jorge Andre Swieca Summer School on
  Particle and Fields}}}}\ (\bibinfo {year} {2005})\ \Eprint
  {https://arxiv.org/abs/hep-th/0504095} {arXiv:hep-th/0504095} \BibitemShut
  {NoStop}%
\bibitem [{\citenamefont {Dudal}\ \emph
  {et~al.}(2008{\natexlab{a}})\citenamefont {Dudal}, \citenamefont {Sorella},
  \citenamefont {Vandersickel},\ and\ \citenamefont
  {Verschelde}}]{Dudal:2007cw}%
  \BibitemOpen
  \bibfield  {author} {\bibinfo {author} {\bibfnamefont {D.}~\bibnamefont
  {Dudal}}, \bibinfo {author} {\bibfnamefont {S.~P.}\ \bibnamefont {Sorella}},
  \bibinfo {author} {\bibfnamefont {N.}~\bibnamefont {Vandersickel}},\ and\
  \bibinfo {author} {\bibfnamefont {H.}~\bibnamefont {Verschelde}},\ }\bibfield
   {title} {\bibinfo {title} {{New features of the gluon and ghost propagator
  in the infrared region from the Gribov-Zwanziger approach}},\ }\href
  {https://doi.org/10.1103/PhysRevD.77.071501} {\bibfield  {journal} {\bibinfo
  {journal} {Phys. Rev.}\ }\textbf {\bibinfo {volume} {D77}},\ \bibinfo {pages}
  {071501} (\bibinfo {year} {2008}{\natexlab{a}})},\ \Eprint
  {https://arxiv.org/abs/0711.4496} {arXiv:0711.4496 [hep-th]} \BibitemShut
  {NoStop}%
\bibitem [{\citenamefont {Dudal}\ \emph
  {et~al.}(2008{\natexlab{b}})\citenamefont {Dudal}, \citenamefont {Gracey},
  \citenamefont {Sorella}, \citenamefont {Vandersickel},\ and\ \citenamefont
  {Verschelde}}]{Dudal:2008sp}%
  \BibitemOpen
  \bibfield  {author} {\bibinfo {author} {\bibfnamefont {D.}~\bibnamefont
  {Dudal}}, \bibinfo {author} {\bibfnamefont {J.~A.}\ \bibnamefont {Gracey}},
  \bibinfo {author} {\bibfnamefont {S.~P.}\ \bibnamefont {Sorella}}, \bibinfo
  {author} {\bibfnamefont {N.}~\bibnamefont {Vandersickel}},\ and\ \bibinfo
  {author} {\bibfnamefont {H.}~\bibnamefont {Verschelde}},\ }\bibfield  {title}
  {\bibinfo {title} {{A Refinement of the Gribov-Zwanziger approach in the
  Landau gauge: Infrared propagators in harmony with the lattice results}},\
  }\href {https://doi.org/10.1103/PhysRevD.78.065047} {\bibfield  {journal}
  {\bibinfo  {journal} {Phys. Rev.}\ }\textbf {\bibinfo {volume} {D78}},\
  \bibinfo {pages} {065047} (\bibinfo {year} {2008}{\natexlab{b}})},\ \Eprint
  {https://arxiv.org/abs/0806.4348} {arXiv:0806.4348 [hep-th]} \BibitemShut
  {NoStop}%
\bibitem [{\citenamefont {Dudal}\ \emph {et~al.}(2010)\citenamefont {Dudal},
  \citenamefont {Oliveira},\ and\ \citenamefont {Vandersickel}}]{Dudal:2010tf}%
  \BibitemOpen
  \bibfield  {author} {\bibinfo {author} {\bibfnamefont {D.}~\bibnamefont
  {Dudal}}, \bibinfo {author} {\bibfnamefont {O.}~\bibnamefont {Oliveira}},\
  and\ \bibinfo {author} {\bibfnamefont {N.}~\bibnamefont {Vandersickel}},\
  }\bibfield  {title} {\bibinfo {title} {{Indirect lattice evidence for the
  Refined Gribov-Zwanziger formalism and the gluon condensate
  $\langle{A^2}\rangle$ in the Landau gauge}},\ }\href
  {https://doi.org/10.1103/PhysRevD.81.074505} {\bibfield  {journal} {\bibinfo
  {journal} {Phys. Rev. D}\ }\textbf {\bibinfo {volume} {81}},\ \bibinfo
  {pages} {074505} (\bibinfo {year} {2010})},\ \Eprint
  {https://arxiv.org/abs/1002.2374} {arXiv:1002.2374 [hep-lat]} \BibitemShut
  {NoStop}%
\bibitem [{\citenamefont {Cucchieri}\ \emph {et~al.}(2012)\citenamefont
  {Cucchieri}, \citenamefont {Dudal}, \citenamefont {Mendes},\ and\
  \citenamefont {Vandersickel}}]{Cucchieri:2011ig}%
  \BibitemOpen
  \bibfield  {author} {\bibinfo {author} {\bibfnamefont {A.}~\bibnamefont
  {Cucchieri}}, \bibinfo {author} {\bibfnamefont {D.}~\bibnamefont {Dudal}},
  \bibinfo {author} {\bibfnamefont {T.}~\bibnamefont {Mendes}},\ and\ \bibinfo
  {author} {\bibfnamefont {N.}~\bibnamefont {Vandersickel}},\ }\bibfield
  {title} {\bibinfo {title} {{Modeling the Gluon Propagator in Landau Gauge:
  Lattice Estimates of Pole Masses and Dimension-Two Condensates}},\ }\href
  {https://doi.org/10.1103/PhysRevD.85.094513} {\bibfield  {journal} {\bibinfo
  {journal} {Phys. Rev. D}\ }\textbf {\bibinfo {volume} {85}},\ \bibinfo
  {pages} {094513} (\bibinfo {year} {2012})},\ \Eprint
  {https://arxiv.org/abs/1111.2327} {arXiv:1111.2327 [hep-lat]} \BibitemShut
  {NoStop}%
\bibitem [{\citenamefont {Dudal}\ \emph {et~al.}(2018)\citenamefont {Dudal},
  \citenamefont {Oliveira},\ and\ \citenamefont {Silva}}]{Dudal:2018cli}%
  \BibitemOpen
  \bibfield  {author} {\bibinfo {author} {\bibfnamefont {D.}~\bibnamefont
  {Dudal}}, \bibinfo {author} {\bibfnamefont {O.}~\bibnamefont {Oliveira}},\
  and\ \bibinfo {author} {\bibfnamefont {P.~J.}\ \bibnamefont {Silva}},\
  }\bibfield  {title} {\bibinfo {title} {{High precision statistical Landau
  gauge lattice gluon propagator computation vs.~the
  Gribov\textendash{}Zwanziger approach}},\ }\href
  {https://doi.org/10.1016/j.aop.2018.08.019} {\bibfield  {journal} {\bibinfo
  {journal} {Annals Phys.}\ }\textbf {\bibinfo {volume} {397}},\ \bibinfo
  {pages} {351} (\bibinfo {year} {2018})},\ \Eprint
  {https://arxiv.org/abs/1803.02281} {arXiv:1803.02281 [hep-lat]} \BibitemShut
  {NoStop}%
\bibitem [{\citenamefont {Dudal}\ \emph {et~al.}(2005)\citenamefont {Dudal},
  \citenamefont {Sobreiro}, \citenamefont {Sorella},\ and\ \citenamefont
  {Verschelde}}]{Dudal:2005na}%
  \BibitemOpen
  \bibfield  {author} {\bibinfo {author} {\bibfnamefont {D.}~\bibnamefont
  {Dudal}}, \bibinfo {author} {\bibfnamefont {R.~F.}\ \bibnamefont {Sobreiro}},
  \bibinfo {author} {\bibfnamefont {S.~P.}\ \bibnamefont {Sorella}},\ and\
  \bibinfo {author} {\bibfnamefont {H.}~\bibnamefont {Verschelde}},\ }\bibfield
   {title} {\bibinfo {title} {{The Gribov parameter and the dimension two gluon
  condensate in Euclidean Yang-Mills theories in the Landau gauge}},\ }\href
  {https://doi.org/10.1103/PhysRevD.72.014016} {\bibfield  {journal} {\bibinfo
  {journal} {Phys. Rev. D}\ }\textbf {\bibinfo {volume} {72}},\ \bibinfo
  {pages} {014016} (\bibinfo {year} {2005})},\ \Eprint
  {https://arxiv.org/abs/hep-th/0502183} {arXiv:hep-th/0502183} \BibitemShut
  {NoStop}%
\bibitem [{\citenamefont {Dudal}\ \emph {et~al.}(2011)\citenamefont {Dudal},
  \citenamefont {Sorella},\ and\ \citenamefont {Vandersickel}}]{Dudal:2011gd}%
  \BibitemOpen
  \bibfield  {author} {\bibinfo {author} {\bibfnamefont {D.}~\bibnamefont
  {Dudal}}, \bibinfo {author} {\bibfnamefont {S.~P.}\ \bibnamefont {Sorella}},\
  and\ \bibinfo {author} {\bibfnamefont {N.}~\bibnamefont {Vandersickel}},\
  }\bibfield  {title} {\bibinfo {title} {{The dynamical origin of the
  refinement of the Gribov-Zwanziger theory}},\ }\href
  {https://doi.org/10.1103/PhysRevD.84.065039} {\bibfield  {journal} {\bibinfo
  {journal} {Phys. Rev. D}\ }\textbf {\bibinfo {volume} {84}},\ \bibinfo
  {pages} {065039} (\bibinfo {year} {2011})},\ \Eprint
  {https://arxiv.org/abs/1105.3371} {arXiv:1105.3371 [hep-th]} \BibitemShut
  {NoStop}%
\bibitem [{\citenamefont {Dudal}\ \emph {et~al.}(2019)\citenamefont {Dudal},
  \citenamefont {Felix}, \citenamefont {Palhares}, \citenamefont {Rondeau},\
  and\ \citenamefont {Vercauteren}}]{Dudal:2019ing}%
  \BibitemOpen
  \bibfield  {author} {\bibinfo {author} {\bibfnamefont {D.}~\bibnamefont
  {Dudal}}, \bibinfo {author} {\bibfnamefont {C.~P.}\ \bibnamefont {Felix}},
  \bibinfo {author} {\bibfnamefont {L.~F.}\ \bibnamefont {Palhares}}, \bibinfo
  {author} {\bibfnamefont {F.}~\bibnamefont {Rondeau}},\ and\ \bibinfo {author}
  {\bibfnamefont {D.}~\bibnamefont {Vercauteren}},\ }\bibfield  {title}
  {\bibinfo {title} {{The BRST-invariant vacuum state of the
  Gribov\textendash{}Zwanziger theory}},\ }\href
  {https://doi.org/10.1140/epjc/s10052-019-7235-0} {\bibfield  {journal}
  {\bibinfo  {journal} {Eur. Phys. J. C}\ }\textbf {\bibinfo {volume} {79}},\
  \bibinfo {pages} {731} (\bibinfo {year} {2019})},\ \Eprint
  {https://arxiv.org/abs/1901.11264} {arXiv:1901.11264 [hep-th]} \BibitemShut
  {NoStop}%
\bibitem [{\citenamefont {Tissier}\ and\ \citenamefont
  {Wschebor}(2010)}]{Tissier:2010ts}%
  \BibitemOpen
  \bibfield  {author} {\bibinfo {author} {\bibfnamefont {M.}~\bibnamefont
  {Tissier}}\ and\ \bibinfo {author} {\bibfnamefont {N.}~\bibnamefont
  {Wschebor}},\ }\bibfield  {title} {\bibinfo {title} {{Infrared propagators of
  Yang-Mills theory from perturbation theory}},\ }\href
  {https://doi.org/10.1103/PhysRevD.82.101701} {\bibfield  {journal} {\bibinfo
  {journal} {Phys. Rev. D}\ }\textbf {\bibinfo {volume} {82}},\ \bibinfo
  {pages} {101701} (\bibinfo {year} {2010})},\ \Eprint
  {https://arxiv.org/abs/1004.1607} {arXiv:1004.1607 [hep-ph]} \BibitemShut
  {NoStop}%
\bibitem [{\citenamefont {Tissier}\ and\ \citenamefont
  {Wschebor}(2011)}]{Tissier:2011ey}%
  \BibitemOpen
  \bibfield  {author} {\bibinfo {author} {\bibfnamefont {M.}~\bibnamefont
  {Tissier}}\ and\ \bibinfo {author} {\bibfnamefont {N.}~\bibnamefont
  {Wschebor}},\ }\bibfield  {title} {\bibinfo {title} {{An Infrared Safe
  perturbative approach to Yang-Mills correlators}},\ }\href
  {https://doi.org/10.1103/PhysRevD.84.045018} {\bibfield  {journal} {\bibinfo
  {journal} {Phys. Rev. D}\ }\textbf {\bibinfo {volume} {84}},\ \bibinfo
  {pages} {045018} (\bibinfo {year} {2011})},\ \Eprint
  {https://arxiv.org/abs/1105.2475} {arXiv:1105.2475 [hep-th]} \BibitemShut
  {NoStop}%
\bibitem [{\citenamefont {Pelaez}\ \emph {et~al.}(2013)\citenamefont {Pelaez},
  \citenamefont {Tissier},\ and\ \citenamefont {Wschebor}}]{Pelaez:2013cpa}%
  \BibitemOpen
  \bibfield  {author} {\bibinfo {author} {\bibfnamefont {M.}~\bibnamefont
  {Pelaez}}, \bibinfo {author} {\bibfnamefont {M.}~\bibnamefont {Tissier}},\
  and\ \bibinfo {author} {\bibfnamefont {N.}~\bibnamefont {Wschebor}},\
  }\bibfield  {title} {\bibinfo {title} {{Three-point correlation functions in
  Yang-Mills theory}},\ }\href {https://doi.org/10.1103/PhysRevD.88.125003}
  {\bibfield  {journal} {\bibinfo  {journal} {Phys. Rev. D}\ }\textbf {\bibinfo
  {volume} {88}},\ \bibinfo {pages} {125003} (\bibinfo {year} {2013})},\
  \Eprint {https://arxiv.org/abs/1310.2594} {arXiv:1310.2594 [hep-th]}
  \BibitemShut {NoStop}%
\bibitem [{\citenamefont {Reinosa}\ \emph {et~al.}(2017)\citenamefont
  {Reinosa}, \citenamefont {Serreau}, \citenamefont {Tissier},\ and\
  \citenamefont {Wschebor}}]{Reinosa:2017qtf}%
  \BibitemOpen
  \bibfield  {author} {\bibinfo {author} {\bibfnamefont {U.}~\bibnamefont
  {Reinosa}}, \bibinfo {author} {\bibfnamefont {J.}~\bibnamefont {Serreau}},
  \bibinfo {author} {\bibfnamefont {M.}~\bibnamefont {Tissier}},\ and\ \bibinfo
  {author} {\bibfnamefont {N.}~\bibnamefont {Wschebor}},\ }\bibfield  {title}
  {\bibinfo {title} {{How nonperturbative is the infrared regime of Landau
  gauge Yang-Mills correlators?}},\ }\href
  {https://doi.org/10.1103/PhysRevD.96.014005} {\bibfield  {journal} {\bibinfo
  {journal} {Phys. Rev. D}\ }\textbf {\bibinfo {volume} {96}},\ \bibinfo
  {pages} {014005} (\bibinfo {year} {2017})},\ \Eprint
  {https://arxiv.org/abs/1703.04041} {arXiv:1703.04041 [hep-th]} \BibitemShut
  {NoStop}%
\bibitem [{\citenamefont {Barrios}\ \emph
  {et~al.}(2024{\natexlab{a}})\citenamefont {Barrios}, \citenamefont
  {De~Fabritiis},\ and\ \citenamefont {Pel\'aez}}]{Barrios:2024ixj}%
  \BibitemOpen
  \bibfield  {author} {\bibinfo {author} {\bibfnamefont {N.}~\bibnamefont
  {Barrios}}, \bibinfo {author} {\bibfnamefont {P.}~\bibnamefont
  {De~Fabritiis}},\ and\ \bibinfo {author} {\bibfnamefont {M.}~\bibnamefont
  {Pel\'aez}},\ }\bibfield  {title} {\bibinfo {title} {{Four-gluon vertex from
  the Curci-Ferrari model at one-loop order}},\ }\href@noop {} {\  (\bibinfo
  {year} {2024}{\natexlab{a}})},\ \Eprint {https://arxiv.org/abs/2403.17056}
  {arXiv:2403.17056 [hep-th]} \BibitemShut {NoStop}%
\bibitem [{\citenamefont {Gracey}\ \emph {et~al.}(2019)\citenamefont {Gracey},
  \citenamefont {Pel\'aez}, \citenamefont {Reinosa},\ and\ \citenamefont
  {Tissier}}]{Gracey:2019xom}%
  \BibitemOpen
  \bibfield  {author} {\bibinfo {author} {\bibfnamefont {J.~A.}\ \bibnamefont
  {Gracey}}, \bibinfo {author} {\bibfnamefont {M.}~\bibnamefont {Pel\'aez}},
  \bibinfo {author} {\bibfnamefont {U.}~\bibnamefont {Reinosa}},\ and\ \bibinfo
  {author} {\bibfnamefont {M.}~\bibnamefont {Tissier}},\ }\bibfield  {title}
  {\bibinfo {title} {{Two loop calculation of Yang-Mills propagators in the
  Curci-Ferrari model}},\ }\href {https://doi.org/10.1103/PhysRevD.100.034023}
  {\bibfield  {journal} {\bibinfo  {journal} {Phys. Rev. D}\ }\textbf {\bibinfo
  {volume} {100}},\ \bibinfo {pages} {034023} (\bibinfo {year} {2019})},\
  \Eprint {https://arxiv.org/abs/1905.07262} {arXiv:1905.07262 [hep-th]}
  \BibitemShut {NoStop}%
\bibitem [{\citenamefont {Barrios}\ \emph {et~al.}(2020)\citenamefont
  {Barrios}, \citenamefont {Pel\'aez}, \citenamefont {Reinosa},\ and\
  \citenamefont {Wschebor}}]{Barrios:2020ubx}%
  \BibitemOpen
  \bibfield  {author} {\bibinfo {author} {\bibfnamefont {N.}~\bibnamefont
  {Barrios}}, \bibinfo {author} {\bibfnamefont {M.}~\bibnamefont {Pel\'aez}},
  \bibinfo {author} {\bibfnamefont {U.}~\bibnamefont {Reinosa}},\ and\ \bibinfo
  {author} {\bibfnamefont {N.}~\bibnamefont {Wschebor}},\ }\bibfield  {title}
  {\bibinfo {title} {{The ghost-antighost-gluon vertex from the Curci-Ferrari
  model: Two-loop corrections}},\ }\href
  {https://doi.org/10.1103/PhysRevD.102.114016} {\bibfield  {journal} {\bibinfo
   {journal} {Phys. Rev. D}\ }\textbf {\bibinfo {volume} {102}},\ \bibinfo
  {pages} {114016} (\bibinfo {year} {2020})},\ \Eprint
  {https://arxiv.org/abs/2009.00875} {arXiv:2009.00875 [hep-th]} \BibitemShut
  {NoStop}%
\bibitem [{\citenamefont {Gracey}(2012)}]{Gracey:2012wf}%
  \BibitemOpen
  \bibfield  {author} {\bibinfo {author} {\bibfnamefont {J.~A.}\ \bibnamefont
  {Gracey}},\ }\bibfield  {title} {\bibinfo {title} {{Power corrections to
  symmetric point vertices in Gribov-Zwanziger theory}},\ }\href
  {https://doi.org/10.1103/PhysRevD.86.105029} {\bibfield  {journal} {\bibinfo
  {journal} {Phys. Rev. D}\ }\textbf {\bibinfo {volume} {86}},\ \bibinfo
  {pages} {105029} (\bibinfo {year} {2012})},\ \Eprint
  {https://arxiv.org/abs/1210.5962} {arXiv:1210.5962 [hep-th]} \BibitemShut
  {NoStop}%
\bibitem [{\citenamefont {Mintz}\ \emph {et~al.}(2018)\citenamefont {Mintz},
  \citenamefont {Palhares}, \citenamefont {Sorella},\ and\ \citenamefont
  {Pereira}}]{Mintz:2017qri}%
  \BibitemOpen
  \bibfield  {author} {\bibinfo {author} {\bibfnamefont {B.~W.}\ \bibnamefont
  {Mintz}}, \bibinfo {author} {\bibfnamefont {L.~F.}\ \bibnamefont {Palhares}},
  \bibinfo {author} {\bibfnamefont {S.~P.}\ \bibnamefont {Sorella}},\ and\
  \bibinfo {author} {\bibfnamefont {A.~D.}\ \bibnamefont {Pereira}},\
  }\bibfield  {title} {\bibinfo {title} {{Ghost-gluon vertex in the presence of
  the Gribov horizon}},\ }\href {https://doi.org/10.1103/PhysRevD.97.034020}
  {\bibfield  {journal} {\bibinfo  {journal} {Phys. Rev. D}\ }\textbf {\bibinfo
  {volume} {97}},\ \bibinfo {pages} {034020} (\bibinfo {year} {2018})},\
  \Eprint {https://arxiv.org/abs/1712.09633} {arXiv:1712.09633 [hep-th]}
  \BibitemShut {NoStop}%
\bibitem [{\citenamefont {Barrios}\ \emph
  {et~al.}(2024{\natexlab{b}})\citenamefont {Barrios}, \citenamefont
  {Pel\'aez}, \citenamefont {Guimaraes}, \citenamefont {Mintz},\ and\
  \citenamefont {Palhares}}]{Barrios:2024idr}%
  \BibitemOpen
  \bibfield  {author} {\bibinfo {author} {\bibfnamefont {N.}~\bibnamefont
  {Barrios}}, \bibinfo {author} {\bibfnamefont {M.}~\bibnamefont {Pel\'aez}},
  \bibinfo {author} {\bibfnamefont {M.}~\bibnamefont {Guimaraes}}, \bibinfo
  {author} {\bibfnamefont {B.}~\bibnamefont {Mintz}},\ and\ \bibinfo {author}
  {\bibfnamefont {L.}~\bibnamefont {Palhares}},\ }\bibfield  {title} {\bibinfo
  {title} {{The ghost-gluon vertex in the presence of the Gribov horizon:
  general kinematics}},\ }\href@noop {} {\  (\bibinfo {year}
  {2024}{\natexlab{b}})},\ \Eprint {https://arxiv.org/abs/2402.17534}
  {arXiv:2402.17534 [hep-ph]} \BibitemShut {NoStop}%
\bibitem [{\citenamefont {de~Brito}\ \emph {et~al.}(2023)\citenamefont
  {de~Brito}, \citenamefont {De~Fabritiis},\ and\ \citenamefont
  {Pereira}}]{deBrito:2023qfs}%
  \BibitemOpen
  \bibfield  {author} {\bibinfo {author} {\bibfnamefont {G.~P.}\ \bibnamefont
  {de~Brito}}, \bibinfo {author} {\bibfnamefont {P.}~\bibnamefont
  {De~Fabritiis}},\ and\ \bibinfo {author} {\bibfnamefont {A.~D.}\ \bibnamefont
  {Pereira}},\ }\bibfield  {title} {\bibinfo {title} {{Refined Gribov-Zwanziger
  theory coupled to scalar fields in the Landau gauge}},\ }\href
  {https://doi.org/10.1103/PhysRevD.107.114006} {\bibfield  {journal} {\bibinfo
   {journal} {Phys. Rev. D}\ }\textbf {\bibinfo {volume} {107}},\ \bibinfo
  {pages} {114006} (\bibinfo {year} {2023})},\ \Eprint
  {https://arxiv.org/abs/2302.04827} {arXiv:2302.04827 [hep-th]} \BibitemShut
  {NoStop}%
\bibitem [{\citenamefont {Guimaraes}\ and\ \citenamefont
  {Sorella}(2011)}]{Guimaraes:2011sf}%
  \BibitemOpen
  \bibfield  {author} {\bibinfo {author} {\bibfnamefont {M.~S.}\ \bibnamefont
  {Guimaraes}}\ and\ \bibinfo {author} {\bibfnamefont {S.~P.}\ \bibnamefont
  {Sorella}},\ }\bibfield  {title} {\bibinfo {title} {{A Few remarks on the
  zero modes of the Faddeev-Popov operator in the Landau and maximal Abelian
  gauges}},\ }\href {https://doi.org/10.1063/1.3641892} {\bibfield  {journal}
  {\bibinfo  {journal} {J. Math. Phys.}\ }\textbf {\bibinfo {volume} {52}},\
  \bibinfo {pages} {092302} (\bibinfo {year} {2011})},\ \Eprint
  {https://arxiv.org/abs/1106.3944} {arXiv:1106.3944 [hep-th]} \BibitemShut
  {NoStop}%
\bibitem [{\citenamefont {van Baal}(1992)}]{vanBaal:1991zw}%
  \BibitemOpen
  \bibfield  {author} {\bibinfo {author} {\bibfnamefont {P.}~\bibnamefont {van
  Baal}},\ }\bibfield  {title} {\bibinfo {title} {{More (thoughts on) Gribov
  copies}},\ }\href {https://doi.org/10.1016/0550-3213(92)90386-P} {\bibfield
  {journal} {\bibinfo  {journal} {Nucl. Phys.}\ }\textbf {\bibinfo {volume}
  {B369}},\ \bibinfo {pages} {259} (\bibinfo {year} {1992})}\BibitemShut
  {NoStop}%
\bibitem [{\citenamefont {Singer}(1978)}]{Singer:1978dk}%
  \BibitemOpen
  \bibfield  {author} {\bibinfo {author} {\bibfnamefont {I.}~\bibnamefont
  {Singer}},\ }\bibfield  {title} {\bibinfo {title} {{Some Remarks on the
  Gribov Ambiguity}},\ }\href {https://doi.org/10.1007/BF01609471} {\bibfield
  {journal} {\bibinfo  {journal} {Commun. Math. Phys.}\ }\textbf {\bibinfo
  {volume} {60}},\ \bibinfo {pages} {7} (\bibinfo {year} {1978})}\BibitemShut
  {NoStop}%
\bibitem [{\citenamefont {Zwanziger}(1989{\natexlab{a}})}]{Zwanziger:1989mf}%
  \BibitemOpen
  \bibfield  {author} {\bibinfo {author} {\bibfnamefont {D.}~\bibnamefont
  {Zwanziger}},\ }\bibfield  {title} {\bibinfo {title} {{Local and
  Renormalizable Action From the Gribov Horizon}},\ }\href
  {https://doi.org/10.1016/0550-3213(89)90122-3} {\bibfield  {journal}
  {\bibinfo  {journal} {Nucl. Phys.}\ }\textbf {\bibinfo {volume} {B323}},\
  \bibinfo {pages} {513} (\bibinfo {year} {1989}{\natexlab{a}})}\BibitemShut
  {NoStop}%
\bibitem [{\citenamefont {Zwanziger}(1989{\natexlab{b}})}]{Zwanziger:1988jt}%
  \BibitemOpen
  \bibfield  {author} {\bibinfo {author} {\bibfnamefont {D.}~\bibnamefont
  {Zwanziger}},\ }\bibfield  {title} {\bibinfo {title} {{Action From the Gribov
  Horizon}},\ }\href {https://doi.org/10.1016/0550-3213(89)90263-0} {\bibfield
  {journal} {\bibinfo  {journal} {Nucl. Phys. B}\ }\textbf {\bibinfo {volume}
  {321}},\ \bibinfo {pages} {591} (\bibinfo {year}
  {1989}{\natexlab{b}})}\BibitemShut {NoStop}%
\bibitem [{\citenamefont {Capri}\ \emph {et~al.}(2013)\citenamefont {Capri},
  \citenamefont {Dudal}, \citenamefont {Guimaraes}, \citenamefont {Palhares},\
  and\ \citenamefont {Sorella}}]{Capri:2012wx}%
  \BibitemOpen
  \bibfield  {author} {\bibinfo {author} {\bibfnamefont {M.~A.~L.}\
  \bibnamefont {Capri}}, \bibinfo {author} {\bibfnamefont {D.}~\bibnamefont
  {Dudal}}, \bibinfo {author} {\bibfnamefont {M.~S.}\ \bibnamefont
  {Guimaraes}}, \bibinfo {author} {\bibfnamefont {L.~F.}\ \bibnamefont
  {Palhares}},\ and\ \bibinfo {author} {\bibfnamefont {S.~P.}\ \bibnamefont
  {Sorella}},\ }\bibfield  {title} {\bibinfo {title} {{An all-order proof of
  the equivalence between Gribov's no-pole and Zwanziger's horizon
  conditions}},\ }\href {https://doi.org/10.1016/j.physletb.2013.01.039}
  {\bibfield  {journal} {\bibinfo  {journal} {Phys. Lett.}\ }\textbf {\bibinfo
  {volume} {B719}},\ \bibinfo {pages} {448} (\bibinfo {year} {2013})},\ \Eprint
  {https://arxiv.org/abs/1212.2419} {arXiv:1212.2419 [hep-th]} \BibitemShut
  {NoStop}%
\bibitem [{\citenamefont {Dell'Antonio}\ and\ \citenamefont
  {Zwanziger}(1991)}]{DellAntonio:1991mms}%
  \BibitemOpen
  \bibfield  {author} {\bibinfo {author} {\bibfnamefont {G.}~\bibnamefont
  {Dell'Antonio}}\ and\ \bibinfo {author} {\bibfnamefont {D.}~\bibnamefont
  {Zwanziger}},\ }\bibfield  {title} {\bibinfo {title} {{Every gauge orbit
  passes inside the Gribov horizon}},\ }\href
  {https://doi.org/10.1007/BF02099494} {\bibfield  {journal} {\bibinfo
  {journal} {Commun. Math. Phys.}\ }\textbf {\bibinfo {volume} {138}},\
  \bibinfo {pages} {291} (\bibinfo {year} {1991})}\BibitemShut {NoStop}%
\bibitem [{\citenamefont {Kugo}\ and\ \citenamefont
  {Ojima}(1979)}]{Kugo:1979gm}%
  \BibitemOpen
  \bibfield  {author} {\bibinfo {author} {\bibfnamefont {T.}~\bibnamefont
  {Kugo}}\ and\ \bibinfo {author} {\bibfnamefont {I.}~\bibnamefont {Ojima}},\
  }\bibfield  {title} {\bibinfo {title} {{Local Covariant Operator Formalism of
  Nonabelian Gauge Theories and Quark Confinement Problem}},\ }\href
  {https://doi.org/10.1143/PTPS.66.1} {\bibfield  {journal} {\bibinfo
  {journal} {Prog. Theor. Phys. Suppl.}\ }\textbf {\bibinfo {volume} {66}},\
  \bibinfo {pages} {1} (\bibinfo {year} {1979})}\BibitemShut {NoStop}%
\bibitem [{\citenamefont {von Smekal}\ \emph {et~al.}(1997)\citenamefont {von
  Smekal}, \citenamefont {Alkofer},\ and\ \citenamefont
  {Hauck}}]{vonSmekal:1997ohs}%
  \BibitemOpen
  \bibfield  {author} {\bibinfo {author} {\bibfnamefont {L.}~\bibnamefont {von
  Smekal}}, \bibinfo {author} {\bibfnamefont {R.}~\bibnamefont {Alkofer}},\
  and\ \bibinfo {author} {\bibfnamefont {A.}~\bibnamefont {Hauck}},\ }\bibfield
   {title} {\bibinfo {title} {{The Infrared behavior of gluon and ghost
  propagators in Landau gauge QCD}},\ }\href
  {https://doi.org/10.1103/PhysRevLett.79.3591} {\bibfield  {journal} {\bibinfo
   {journal} {Phys. Rev. Lett.}\ }\textbf {\bibinfo {volume} {79}},\ \bibinfo
  {pages} {3591} (\bibinfo {year} {1997})},\ \Eprint
  {https://arxiv.org/abs/hep-ph/9705242} {arXiv:hep-ph/9705242} \BibitemShut
  {NoStop}%
\bibitem [{\citenamefont {von Smekal}\ \emph {et~al.}(1998)\citenamefont {von
  Smekal}, \citenamefont {Hauck},\ and\ \citenamefont
  {Alkofer}}]{vonSmekal:1997ern}%
  \BibitemOpen
  \bibfield  {author} {\bibinfo {author} {\bibfnamefont {L.}~\bibnamefont {von
  Smekal}}, \bibinfo {author} {\bibfnamefont {A.}~\bibnamefont {Hauck}},\ and\
  \bibinfo {author} {\bibfnamefont {R.}~\bibnamefont {Alkofer}},\ }\bibfield
  {title} {\bibinfo {title} {{A Solution to Coupled Dyson\textendash{}Schwinger
  Equations for Gluons and Ghosts in Landau Gauge}},\ }\href
  {https://doi.org/10.1006/aphy.1998.5806} {\bibfield  {journal} {\bibinfo
  {journal} {Annals Phys.}\ }\textbf {\bibinfo {volume} {267}},\ \bibinfo
  {pages} {1} (\bibinfo {year} {1998})},\ \bibinfo {note} {[Erratum: Annals
  Phys. 269, 182 (1998)]},\ \Eprint {https://arxiv.org/abs/hep-ph/9707327}
  {arXiv:hep-ph/9707327} \BibitemShut {NoStop}%
\bibitem [{\citenamefont {Zwanziger}(2002)}]{Zwanziger:2001kw}%
  \BibitemOpen
  \bibfield  {author} {\bibinfo {author} {\bibfnamefont {D.}~\bibnamefont
  {Zwanziger}},\ }\bibfield  {title} {\bibinfo {title} {{Nonperturbative Landau
  gauge and infrared critical exponents in QCD}},\ }\href
  {https://doi.org/10.1103/PhysRevD.65.094039} {\bibfield  {journal} {\bibinfo
  {journal} {Phys. Rev. D}\ }\textbf {\bibinfo {volume} {65}},\ \bibinfo
  {pages} {094039} (\bibinfo {year} {2002})},\ \Eprint
  {https://arxiv.org/abs/hep-th/0109224} {arXiv:hep-th/0109224} \BibitemShut
  {NoStop}%
\bibitem [{\citenamefont {Gracey}(2006)}]{Gracey:2006dr}%
  \BibitemOpen
  \bibfield  {author} {\bibinfo {author} {\bibfnamefont {J.~A.}\ \bibnamefont
  {Gracey}},\ }\bibfield  {title} {\bibinfo {title} {{One loop gluon form
  factor and freezing of alpha(s) in the Gribov-Zwanziger QCD Lagrangian}},\
  }\href {https://doi.org/10.1088/1126-6708/2006/05/052} {\bibfield  {journal}
  {\bibinfo  {journal} {JHEP}\ }\textbf {\bibinfo {volume} {05}},\ \bibinfo
  {pages} {052}},\ \bibinfo {note} {[Erratum: JHEP 02, 078 (2010)]},\ \Eprint
  {https://arxiv.org/abs/hep-ph/0605077} {arXiv:hep-ph/0605077} \BibitemShut
  {NoStop}%
\bibitem [{\citenamefont {Huber}\ \emph {et~al.}(2010)\citenamefont {Huber},
  \citenamefont {Alkofer},\ and\ \citenamefont {Sorella}}]{Huber:2009tx}%
  \BibitemOpen
  \bibfield  {author} {\bibinfo {author} {\bibfnamefont {M.~Q.}\ \bibnamefont
  {Huber}}, \bibinfo {author} {\bibfnamefont {R.}~\bibnamefont {Alkofer}},\
  and\ \bibinfo {author} {\bibfnamefont {S.~P.}\ \bibnamefont {Sorella}},\
  }\bibfield  {title} {\bibinfo {title} {{Infrared analysis of Dyson-Schwinger
  equations taking into account the Gribov horizon in Landau gauge}},\ }\href
  {https://doi.org/10.1103/PhysRevD.81.065003} {\bibfield  {journal} {\bibinfo
  {journal} {Phys. Rev. D}\ }\textbf {\bibinfo {volume} {81}},\ \bibinfo
  {pages} {065003} (\bibinfo {year} {2010})},\ \Eprint
  {https://arxiv.org/abs/0910.5604} {arXiv:0910.5604 [hep-th]} \BibitemShut
  {NoStop}%
\bibitem [{\citenamefont {Cucchieri}\ and\ \citenamefont
  {Mendes}(2008{\natexlab{a}})}]{Cucchieri:2007rg}%
  \BibitemOpen
  \bibfield  {author} {\bibinfo {author} {\bibfnamefont {A.}~\bibnamefont
  {Cucchieri}}\ and\ \bibinfo {author} {\bibfnamefont {T.}~\bibnamefont
  {Mendes}},\ }\bibfield  {title} {\bibinfo {title} {{Constraints on the IR
  behavior of the gluon propagator in Yang-Mills theories}},\ }\href
  {https://doi.org/10.1103/PhysRevLett.100.241601} {\bibfield  {journal}
  {\bibinfo  {journal} {Phys. Rev. Lett.}\ }\textbf {\bibinfo {volume} {100}},\
  \bibinfo {pages} {241601} (\bibinfo {year} {2008}{\natexlab{a}})},\ \Eprint
  {https://arxiv.org/abs/0712.3517} {arXiv:0712.3517 [hep-lat]} \BibitemShut
  {NoStop}%
\bibitem [{\citenamefont {Cucchieri}\ and\ \citenamefont
  {Mendes}(2008{\natexlab{b}})}]{Cucchieri:2008fc}%
  \BibitemOpen
  \bibfield  {author} {\bibinfo {author} {\bibfnamefont {A.}~\bibnamefont
  {Cucchieri}}\ and\ \bibinfo {author} {\bibfnamefont {T.}~\bibnamefont
  {Mendes}},\ }\bibfield  {title} {\bibinfo {title} {{Constraints on the IR
  behavior of the ghost propagator in Yang-Mills theories}},\ }\href
  {https://doi.org/10.1103/PhysRevD.78.094503} {\bibfield  {journal} {\bibinfo
  {journal} {Phys. Rev. D}\ }\textbf {\bibinfo {volume} {78}},\ \bibinfo
  {pages} {094503} (\bibinfo {year} {2008}{\natexlab{b}})},\ \Eprint
  {https://arxiv.org/abs/0804.2371} {arXiv:0804.2371 [hep-lat]} \BibitemShut
  {NoStop}%
\bibitem [{\citenamefont {Bornyakov}\ \emph {et~al.}(2009)\citenamefont
  {Bornyakov}, \citenamefont {Mitrjushkin},\ and\ \citenamefont
  {Muller-Preussker}}]{Bornyakov:2008yx}%
  \BibitemOpen
  \bibfield  {author} {\bibinfo {author} {\bibfnamefont {V.~G.}\ \bibnamefont
  {Bornyakov}}, \bibinfo {author} {\bibfnamefont {V.~K.}\ \bibnamefont
  {Mitrjushkin}},\ and\ \bibinfo {author} {\bibfnamefont {M.}~\bibnamefont
  {Muller-Preussker}},\ }\bibfield  {title} {\bibinfo {title} {{Infrared
  behavior and Gribov ambiguity in SU(2) lattice gauge theory}},\ }\href
  {https://doi.org/10.1103/PhysRevD.79.074504} {\bibfield  {journal} {\bibinfo
  {journal} {Phys. Rev. D}\ }\textbf {\bibinfo {volume} {79}},\ \bibinfo
  {pages} {074504} (\bibinfo {year} {2009})},\ \Eprint
  {https://arxiv.org/abs/0812.2761} {arXiv:0812.2761 [hep-lat]} \BibitemShut
  {NoStop}%
\bibitem [{\citenamefont {Bogolubsky}\ \emph {et~al.}(2009)\citenamefont
  {Bogolubsky}, \citenamefont {Ilgenfritz}, \citenamefont {Muller-Preussker},\
  and\ \citenamefont {Sternbeck}}]{Bogolubsky:2009dc}%
  \BibitemOpen
  \bibfield  {author} {\bibinfo {author} {\bibfnamefont {I.~L.}\ \bibnamefont
  {Bogolubsky}}, \bibinfo {author} {\bibfnamefont {E.~M.}\ \bibnamefont
  {Ilgenfritz}}, \bibinfo {author} {\bibfnamefont {M.}~\bibnamefont
  {Muller-Preussker}},\ and\ \bibinfo {author} {\bibfnamefont {A.}~\bibnamefont
  {Sternbeck}},\ }\bibfield  {title} {\bibinfo {title} {{Lattice gluodynamics
  computation of Landau gauge Green's functions in the deep infrared}},\ }\href
  {https://doi.org/10.1016/j.physletb.2009.04.076} {\bibfield  {journal}
  {\bibinfo  {journal} {Phys. Lett.}\ }\textbf {\bibinfo {volume} {B676}},\
  \bibinfo {pages} {69} (\bibinfo {year} {2009})},\ \Eprint
  {https://arxiv.org/abs/0901.0736} {arXiv:0901.0736 [hep-lat]} \BibitemShut
  {NoStop}%
\bibitem [{\citenamefont {Aguilar}\ \emph {et~al.}(2008)\citenamefont
  {Aguilar}, \citenamefont {Binosi},\ and\ \citenamefont
  {Papavassiliou}}]{Aguilar:2008xm}%
  \BibitemOpen
  \bibfield  {author} {\bibinfo {author} {\bibfnamefont {A.~C.}\ \bibnamefont
  {Aguilar}}, \bibinfo {author} {\bibfnamefont {D.}~\bibnamefont {Binosi}},\
  and\ \bibinfo {author} {\bibfnamefont {J.}~\bibnamefont {Papavassiliou}},\
  }\bibfield  {title} {\bibinfo {title} {{Gluon and ghost propagators in the
  Landau gauge: Deriving lattice results from Schwinger-Dyson equations}},\
  }\href {https://doi.org/10.1103/PhysRevD.78.025010} {\bibfield  {journal}
  {\bibinfo  {journal} {Phys. Rev. D}\ }\textbf {\bibinfo {volume} {78}},\
  \bibinfo {pages} {025010} (\bibinfo {year} {2008})},\ \Eprint
  {https://arxiv.org/abs/0802.1870} {arXiv:0802.1870 [hep-ph]} \BibitemShut
  {NoStop}%
\bibitem [{\citenamefont {Fischer}\ \emph {et~al.}(2009)\citenamefont
  {Fischer}, \citenamefont {Maas},\ and\ \citenamefont
  {Pawlowski}}]{Fischer:2008uz}%
  \BibitemOpen
  \bibfield  {author} {\bibinfo {author} {\bibfnamefont {C.~S.}\ \bibnamefont
  {Fischer}}, \bibinfo {author} {\bibfnamefont {A.}~\bibnamefont {Maas}},\ and\
  \bibinfo {author} {\bibfnamefont {J.~M.}\ \bibnamefont {Pawlowski}},\
  }\bibfield  {title} {\bibinfo {title} {{On the infrared behavior of Landau
  gauge Yang-Mills theory}},\ }\href
  {https://doi.org/10.1016/j.aop.2009.07.009} {\bibfield  {journal} {\bibinfo
  {journal} {Annals Phys.}\ }\textbf {\bibinfo {volume} {324}},\ \bibinfo
  {pages} {2408} (\bibinfo {year} {2009})},\ \Eprint
  {https://arxiv.org/abs/0810.1987} {arXiv:0810.1987 [hep-ph]} \BibitemShut
  {NoStop}%
\bibitem [{\citenamefont {Alkofer}\ \emph {et~al.}(2010)\citenamefont
  {Alkofer}, \citenamefont {Huber},\ and\ \citenamefont
  {Schwenzer}}]{Alkofer:2008jy}%
  \BibitemOpen
  \bibfield  {author} {\bibinfo {author} {\bibfnamefont {R.}~\bibnamefont
  {Alkofer}}, \bibinfo {author} {\bibfnamefont {M.~Q.}\ \bibnamefont {Huber}},\
  and\ \bibinfo {author} {\bibfnamefont {K.}~\bibnamefont {Schwenzer}},\
  }\bibfield  {title} {\bibinfo {title} {{Infrared singularities in Landau
  gauge Yang-Mills theory}},\ }\href
  {https://doi.org/10.1103/PhysRevD.81.105010} {\bibfield  {journal} {\bibinfo
  {journal} {Phys. Rev. D}\ }\textbf {\bibinfo {volume} {81}},\ \bibinfo
  {pages} {105010} (\bibinfo {year} {2010})},\ \Eprint
  {https://arxiv.org/abs/0801.2762} {arXiv:0801.2762 [hep-th]} \BibitemShut
  {NoStop}%
\bibitem [{\citenamefont {Maggiore}\ and\ \citenamefont
  {Schaden}(1994)}]{Maggiore:1993wq}%
  \BibitemOpen
  \bibfield  {author} {\bibinfo {author} {\bibfnamefont {N.}~\bibnamefont
  {Maggiore}}\ and\ \bibinfo {author} {\bibfnamefont {M.}~\bibnamefont
  {Schaden}},\ }\bibfield  {title} {\bibinfo {title} {{Landau gauge within the
  Gribov horizon}},\ }\href {https://doi.org/10.1103/PhysRevD.50.6616}
  {\bibfield  {journal} {\bibinfo  {journal} {Phys. Rev. D}\ }\textbf {\bibinfo
  {volume} {50}},\ \bibinfo {pages} {6616} (\bibinfo {year} {1994})},\ \Eprint
  {https://arxiv.org/abs/hep-th/9310111} {arXiv:hep-th/9310111} \BibitemShut
  {NoStop}%
\bibitem [{\citenamefont {Baulieu}\ and\ \citenamefont
  {Sorella}(2009)}]{Baulieu:2008fy}%
  \BibitemOpen
  \bibfield  {author} {\bibinfo {author} {\bibfnamefont {L.}~\bibnamefont
  {Baulieu}}\ and\ \bibinfo {author} {\bibfnamefont {S.~P.}\ \bibnamefont
  {Sorella}},\ }\bibfield  {title} {\bibinfo {title} {{Soft breaking of BRST
  invariance for introducing non-perturbative infrared effects in a local and
  renormalizable way}},\ }\href
  {https://doi.org/10.1016/j.physletb.2008.11.036} {\bibfield  {journal}
  {\bibinfo  {journal} {Phys. Lett. B}\ }\textbf {\bibinfo {volume} {671}},\
  \bibinfo {pages} {481} (\bibinfo {year} {2009})},\ \Eprint
  {https://arxiv.org/abs/0808.1356} {arXiv:0808.1356 [hep-th]} \BibitemShut
  {NoStop}%
\bibitem [{\citenamefont {Dudal}\ \emph {et~al.}(2009)\citenamefont {Dudal},
  \citenamefont {Sorella}, \citenamefont {Vandersickel},\ and\ \citenamefont
  {Verschelde}}]{Dudal:2009xh}%
  \BibitemOpen
  \bibfield  {author} {\bibinfo {author} {\bibfnamefont {D.}~\bibnamefont
  {Dudal}}, \bibinfo {author} {\bibfnamefont {S.~P.}\ \bibnamefont {Sorella}},
  \bibinfo {author} {\bibfnamefont {N.}~\bibnamefont {Vandersickel}},\ and\
  \bibinfo {author} {\bibfnamefont {H.}~\bibnamefont {Verschelde}},\ }\bibfield
   {title} {\bibinfo {title} {{Gribov no-pole condition, Zwanziger horizon
  function, Kugo-Ojima confinement criterion, boundary conditions, BRST
  breaking and all that}},\ }\href {https://doi.org/10.1103/PhysRevD.79.121701}
  {\bibfield  {journal} {\bibinfo  {journal} {Phys. Rev. D}\ }\textbf {\bibinfo
  {volume} {79}},\ \bibinfo {pages} {121701} (\bibinfo {year} {2009})},\
  \Eprint {https://arxiv.org/abs/0904.0641} {arXiv:0904.0641 [hep-th]}
  \BibitemShut {NoStop}%
\bibitem [{\citenamefont {Sorella}(2009)}]{Sorella:2009vt}%
  \BibitemOpen
  \bibfield  {author} {\bibinfo {author} {\bibfnamefont {S.~P.}\ \bibnamefont
  {Sorella}},\ }\bibfield  {title} {\bibinfo {title} {{Gribov horizon and BRST
  symmetry: A Few remarks}},\ }\href
  {https://doi.org/10.1103/PhysRevD.80.025013} {\bibfield  {journal} {\bibinfo
  {journal} {Phys. Rev. D}\ }\textbf {\bibinfo {volume} {80}},\ \bibinfo
  {pages} {025013} (\bibinfo {year} {2009})},\ \Eprint
  {https://arxiv.org/abs/0905.1010} {arXiv:0905.1010 [hep-th]} \BibitemShut
  {NoStop}%
\bibitem [{\citenamefont {Sorella}(2011)}]{Sorella:2010it}%
  \BibitemOpen
  \bibfield  {author} {\bibinfo {author} {\bibfnamefont {S.~P.}\ \bibnamefont
  {Sorella}},\ }\bibfield  {title} {\bibinfo {title} {{Gluon confinement,
  i-particles and BRST soft breaking}},\ }\href
  {https://doi.org/10.1088/1751-8113/44/13/135403} {\bibfield  {journal}
  {\bibinfo  {journal} {J. Phys. A}\ }\textbf {\bibinfo {volume} {44}},\
  \bibinfo {pages} {135403} (\bibinfo {year} {2011})},\ \Eprint
  {https://arxiv.org/abs/1006.4500} {arXiv:1006.4500 [hep-th]} \BibitemShut
  {NoStop}%
\bibitem [{\citenamefont {Capri}\ \emph {et~al.}(2010)\citenamefont {Capri},
  \citenamefont {Gomez}, \citenamefont {Guimaraes}, \citenamefont {Lemes},
  \citenamefont {Sorella},\ and\ \citenamefont {Tedesco}}]{Capri:2010hb}%
  \BibitemOpen
  \bibfield  {author} {\bibinfo {author} {\bibfnamefont {M.~A.~L.}\
  \bibnamefont {Capri}}, \bibinfo {author} {\bibfnamefont {A.~J.}\ \bibnamefont
  {Gomez}}, \bibinfo {author} {\bibfnamefont {M.~S.}\ \bibnamefont
  {Guimaraes}}, \bibinfo {author} {\bibfnamefont {V.~E.~R.}\ \bibnamefont
  {Lemes}}, \bibinfo {author} {\bibfnamefont {S.~P.}\ \bibnamefont {Sorella}},\
  and\ \bibinfo {author} {\bibfnamefont {D.~G.}\ \bibnamefont {Tedesco}},\
  }\bibfield  {title} {\bibinfo {title} {{A remark on the BRST symmetry in the
  Gribov-Zwanziger theory}},\ }\href
  {https://doi.org/10.1103/PhysRevD.82.105019} {\bibfield  {journal} {\bibinfo
  {journal} {Phys. Rev. D}\ }\textbf {\bibinfo {volume} {82}},\ \bibinfo
  {pages} {105019} (\bibinfo {year} {2010})},\ \Eprint
  {https://arxiv.org/abs/1009.4135} {arXiv:1009.4135 [hep-th]} \BibitemShut
  {NoStop}%
\bibitem [{\citenamefont {Lavrov}\ \emph {et~al.}(2011)\citenamefont {Lavrov},
  \citenamefont {Lechtenfeld},\ and\ \citenamefont
  {Reshetnyak}}]{Lavrov:2011wb}%
  \BibitemOpen
  \bibfield  {author} {\bibinfo {author} {\bibfnamefont {P.}~\bibnamefont
  {Lavrov}}, \bibinfo {author} {\bibfnamefont {O.}~\bibnamefont
  {Lechtenfeld}},\ and\ \bibinfo {author} {\bibfnamefont {A.}~\bibnamefont
  {Reshetnyak}},\ }\bibfield  {title} {\bibinfo {title} {{Is soft breaking of
  BRST symmetry consistent?}},\ }\href
  {https://doi.org/10.1007/JHEP10(2011)043} {\bibfield  {journal} {\bibinfo
  {journal} {JHEP}\ }\textbf {\bibinfo {volume} {10}},\ \bibinfo {pages}
  {043}},\ \Eprint {https://arxiv.org/abs/1108.4820} {arXiv:1108.4820 [hep-th]}
  \BibitemShut {NoStop}%
\bibitem [{\citenamefont {Serreau}\ and\ \citenamefont
  {Tissier}(2012)}]{Serreau:2012cg}%
  \BibitemOpen
  \bibfield  {author} {\bibinfo {author} {\bibfnamefont {J.}~\bibnamefont
  {Serreau}}\ and\ \bibinfo {author} {\bibfnamefont {M.}~\bibnamefont
  {Tissier}},\ }\bibfield  {title} {\bibinfo {title} {{Lifting the Gribov
  ambiguity in Yang-Mills theories}},\ }\href
  {https://doi.org/10.1016/j.physletb.2012.04.041} {\bibfield  {journal}
  {\bibinfo  {journal} {Phys. Lett. B}\ }\textbf {\bibinfo {volume} {712}},\
  \bibinfo {pages} {97} (\bibinfo {year} {2012})},\ \Eprint
  {https://arxiv.org/abs/1202.3432} {arXiv:1202.3432 [hep-th]} \BibitemShut
  {NoStop}%
\bibitem [{\citenamefont {Serreau}\ \emph {et~al.}(2014)\citenamefont
  {Serreau}, \citenamefont {Tissier},\ and\ \citenamefont
  {Tresmontant}}]{Serreau:2013ila}%
  \BibitemOpen
  \bibfield  {author} {\bibinfo {author} {\bibfnamefont {J.}~\bibnamefont
  {Serreau}}, \bibinfo {author} {\bibfnamefont {M.}~\bibnamefont {Tissier}},\
  and\ \bibinfo {author} {\bibfnamefont {A.}~\bibnamefont {Tresmontant}},\
  }\bibfield  {title} {\bibinfo {title} {{Covariant gauges without Gribov
  ambiguities in Yang-Mills theories}},\ }\href
  {https://doi.org/10.1103/PhysRevD.89.125019} {\bibfield  {journal} {\bibinfo
  {journal} {Phys. Rev. D}\ }\textbf {\bibinfo {volume} {89}},\ \bibinfo
  {pages} {125019} (\bibinfo {year} {2014})},\ \Eprint
  {https://arxiv.org/abs/1307.6019} {arXiv:1307.6019 [hep-th]} \BibitemShut
  {NoStop}%
\bibitem [{\citenamefont {Dudal}\ and\ \citenamefont
  {Sorella}(2012)}]{Dudal:2012sb}%
  \BibitemOpen
  \bibfield  {author} {\bibinfo {author} {\bibfnamefont {D.}~\bibnamefont
  {Dudal}}\ and\ \bibinfo {author} {\bibfnamefont {S.~P.}\ \bibnamefont
  {Sorella}},\ }\bibfield  {title} {\bibinfo {title} {{The Gribov horizon and
  spontaneous BRST symmetry breaking}},\ }\href
  {https://doi.org/10.1103/PhysRevD.86.045005} {\bibfield  {journal} {\bibinfo
  {journal} {Phys. Rev. D}\ }\textbf {\bibinfo {volume} {86}},\ \bibinfo
  {pages} {045005} (\bibinfo {year} {2012})},\ \Eprint
  {https://arxiv.org/abs/1205.3934} {arXiv:1205.3934 [hep-th]} \BibitemShut
  {NoStop}%
\bibitem [{\citenamefont {Pereira}\ and\ \citenamefont
  {Sobreiro}(2013)}]{Pereira:2013aza}%
  \BibitemOpen
  \bibfield  {author} {\bibinfo {author} {\bibfnamefont {A.~D.}\ \bibnamefont
  {Pereira}}\ and\ \bibinfo {author} {\bibfnamefont {R.~F.}\ \bibnamefont
  {Sobreiro}},\ }\bibfield  {title} {\bibinfo {title} {{On the elimination of
  infinitesimal Gribov ambiguities in non-Abelian gauge theories}},\ }\href
  {https://doi.org/10.1140/epjc/s10052-013-2584-6} {\bibfield  {journal}
  {\bibinfo  {journal} {Eur. Phys. J. C}\ }\textbf {\bibinfo {volume} {73}},\
  \bibinfo {pages} {2584} (\bibinfo {year} {2013})},\ \Eprint
  {https://arxiv.org/abs/1308.4159} {arXiv:1308.4159 [hep-th]} \BibitemShut
  {NoStop}%
\bibitem [{\citenamefont {Pereira}\ and\ \citenamefont
  {Sobreiro}(2014)}]{Pereira:2014apa}%
  \BibitemOpen
  \bibfield  {author} {\bibinfo {author} {\bibfnamefont {A.~D.}\ \bibnamefont
  {Pereira}, \bibfnamefont {Jr.}}\ and\ \bibinfo {author} {\bibfnamefont
  {R.~F.}\ \bibnamefont {Sobreiro}},\ }\bibfield  {title} {\bibinfo {title}
  {{Gribov ambiguities at the Landau-maximal Abelian interpolating gauge}},\
  }\href {https://doi.org/10.1140/epjc/s10052-014-2984-2} {\bibfield  {journal}
  {\bibinfo  {journal} {Eur. Phys. J. C}\ }\textbf {\bibinfo {volume} {74}},\
  \bibinfo {pages} {2984} (\bibinfo {year} {2014})},\ \Eprint
  {https://arxiv.org/abs/1402.3477} {arXiv:1402.3477 [hep-th]} \BibitemShut
  {NoStop}%
\bibitem [{\citenamefont {Lavrov}\ and\ \citenamefont
  {Lechtenfeld}(2013)}]{Lavrov:2013boa}%
  \BibitemOpen
  \bibfield  {author} {\bibinfo {author} {\bibfnamefont {P.~M.}\ \bibnamefont
  {Lavrov}}\ and\ \bibinfo {author} {\bibfnamefont {O.}~\bibnamefont
  {Lechtenfeld}},\ }\bibfield  {title} {\bibinfo {title} {{Gribov horizon
  beyond the Landau gauge}},\ }\href
  {https://doi.org/10.1016/j.physletb.2013.07.020} {\bibfield  {journal}
  {\bibinfo  {journal} {Phys. Lett. B}\ }\textbf {\bibinfo {volume} {725}},\
  \bibinfo {pages} {386} (\bibinfo {year} {2013})},\ \Eprint
  {https://arxiv.org/abs/1305.2931} {arXiv:1305.2931 [hep-th]} \BibitemShut
  {NoStop}%
\bibitem [{\citenamefont {Capri}\ \emph {et~al.}(2014)\citenamefont {Capri},
  \citenamefont {Guimaraes}, \citenamefont {Justo}, \citenamefont {Palhares},\
  and\ \citenamefont {Sorella}}]{Capri:2014bsa}%
  \BibitemOpen
  \bibfield  {author} {\bibinfo {author} {\bibfnamefont {M.~A.~L.}\
  \bibnamefont {Capri}}, \bibinfo {author} {\bibfnamefont {M.~S.}\ \bibnamefont
  {Guimaraes}}, \bibinfo {author} {\bibfnamefont {I.~F.}\ \bibnamefont
  {Justo}}, \bibinfo {author} {\bibfnamefont {L.~F.}\ \bibnamefont
  {Palhares}},\ and\ \bibinfo {author} {\bibfnamefont {S.~P.}\ \bibnamefont
  {Sorella}},\ }\bibfield  {title} {\bibinfo {title} {{Properties of the
  Faddeev-Popov operator in the Landau gauge, matter confinement and soft BRST
  breaking}},\ }\href {https://doi.org/10.1103/PhysRevD.90.085010} {\bibfield
  {journal} {\bibinfo  {journal} {Phys. Rev. D}\ }\textbf {\bibinfo {volume}
  {90}},\ \bibinfo {pages} {085010} (\bibinfo {year} {2014})},\ \Eprint
  {https://arxiv.org/abs/1408.3597} {arXiv:1408.3597 [hep-th]} \BibitemShut
  {NoStop}%
\bibitem [{\citenamefont {Cucchieri}\ \emph {et~al.}(2014)\citenamefont
  {Cucchieri}, \citenamefont {Dudal}, \citenamefont {Mendes},\ and\
  \citenamefont {Vandersickel}}]{Cucchieri:2014via}%
  \BibitemOpen
  \bibfield  {author} {\bibinfo {author} {\bibfnamefont {A.}~\bibnamefont
  {Cucchieri}}, \bibinfo {author} {\bibfnamefont {D.}~\bibnamefont {Dudal}},
  \bibinfo {author} {\bibfnamefont {T.}~\bibnamefont {Mendes}},\ and\ \bibinfo
  {author} {\bibfnamefont {N.}~\bibnamefont {Vandersickel}},\ }\bibfield
  {title} {\bibinfo {title} {{BRST-Symmetry Breaking and Bose-Ghost Propagator
  in Lattice Minimal Landau Gauge}},\ }\href
  {https://doi.org/10.1103/PhysRevD.90.051501} {\bibfield  {journal} {\bibinfo
  {journal} {Phys. Rev. D}\ }\textbf {\bibinfo {volume} {90}},\ \bibinfo
  {pages} {051501} (\bibinfo {year} {2014})},\ \Eprint
  {https://arxiv.org/abs/1405.1547} {arXiv:1405.1547 [hep-lat]} \BibitemShut
  {NoStop}%
\bibitem [{\citenamefont {Moshin}\ and\ \citenamefont
  {Reshetnyak}(2014)}]{Moshin:2014xka}%
  \BibitemOpen
  \bibfield  {author} {\bibinfo {author} {\bibfnamefont {P.~Y.}\ \bibnamefont
  {Moshin}}\ and\ \bibinfo {author} {\bibfnamefont {A.~A.}\ \bibnamefont
  {Reshetnyak}},\ }\bibfield  {title} {\bibinfo {title} {{Field-dependent
  BRST\textendash{}antiBRST transformations in Yang\textendash{}Mills and
  Gribov\textendash{}Zwanziger theories}},\ }\href
  {https://doi.org/10.1016/j.nuclphysb.2014.09.011} {\bibfield  {journal}
  {\bibinfo  {journal} {Nucl. Phys. B}\ }\textbf {\bibinfo {volume} {888}},\
  \bibinfo {pages} {92} (\bibinfo {year} {2014})},\ \Eprint
  {https://arxiv.org/abs/1405.0790} {arXiv:1405.0790 [hep-th]} \BibitemShut
  {NoStop}%
\bibitem [{\citenamefont {Schaden}\ and\ \citenamefont
  {Zwanziger}(2015{\natexlab{a}})}]{Schaden:2014bea}%
  \BibitemOpen
  \bibfield  {author} {\bibinfo {author} {\bibfnamefont {M.}~\bibnamefont
  {Schaden}}\ and\ \bibinfo {author} {\bibfnamefont {D.}~\bibnamefont
  {Zwanziger}},\ }\bibfield  {title} {\bibinfo {title} {{Living with
  spontaneously broken BRST symmetry. I. Physical states and cohomology}},\
  }\href {https://doi.org/10.1103/PhysRevD.92.025001} {\bibfield  {journal}
  {\bibinfo  {journal} {Phys. Rev. D}\ }\textbf {\bibinfo {volume} {92}},\
  \bibinfo {pages} {025001} (\bibinfo {year} {2015}{\natexlab{a}})},\ \Eprint
  {https://arxiv.org/abs/1412.4823} {arXiv:1412.4823 [hep-ph]} \BibitemShut
  {NoStop}%
\bibitem [{\citenamefont {Schaden}\ and\ \citenamefont
  {Zwanziger}(2015{\natexlab{b}})}]{Schaden:2015uua}%
  \BibitemOpen
  \bibfield  {author} {\bibinfo {author} {\bibfnamefont {M.}~\bibnamefont
  {Schaden}}\ and\ \bibinfo {author} {\bibfnamefont {D.}~\bibnamefont
  {Zwanziger}},\ }\bibfield  {title} {\bibinfo {title} {{Living with
  spontaneously broken BRST symmetry. II. Poincar\'e invariance}},\ }\href
  {https://doi.org/10.1103/PhysRevD.92.025002} {\bibfield  {journal} {\bibinfo
  {journal} {Phys. Rev. D}\ }\textbf {\bibinfo {volume} {92}},\ \bibinfo
  {pages} {025002} (\bibinfo {year} {2015}{\natexlab{b}})},\ \Eprint
  {https://arxiv.org/abs/1501.05974} {arXiv:1501.05974 [hep-th]} \BibitemShut
  {NoStop}%
\bibitem [{\citenamefont {Capri}\ \emph {et~al.}(2015)\citenamefont {Capri},
  \citenamefont {Dudal}, \citenamefont {Fiorentini}, \citenamefont {Guimaraes},
  \citenamefont {Justo}, \citenamefont {Pereira}, \citenamefont {Mintz},
  \citenamefont {Palhares}, \citenamefont {Sobreiro},\ and\ \citenamefont
  {Sorella}}]{Capri:2015ixa}%
  \BibitemOpen
  \bibfield  {author} {\bibinfo {author} {\bibfnamefont {M.~A.~L.}\
  \bibnamefont {Capri}}, \bibinfo {author} {\bibfnamefont {D.}~\bibnamefont
  {Dudal}}, \bibinfo {author} {\bibfnamefont {D.}~\bibnamefont {Fiorentini}},
  \bibinfo {author} {\bibfnamefont {M.~S.}\ \bibnamefont {Guimaraes}}, \bibinfo
  {author} {\bibfnamefont {I.~F.}\ \bibnamefont {Justo}}, \bibinfo {author}
  {\bibfnamefont {A.~D.}\ \bibnamefont {Pereira}}, \bibinfo {author}
  {\bibfnamefont {B.~W.}\ \bibnamefont {Mintz}}, \bibinfo {author}
  {\bibfnamefont {L.~F.}\ \bibnamefont {Palhares}}, \bibinfo {author}
  {\bibfnamefont {R.~F.}\ \bibnamefont {Sobreiro}},\ and\ \bibinfo {author}
  {\bibfnamefont {S.~P.}\ \bibnamefont {Sorella}},\ }\bibfield  {title}
  {\bibinfo {title} {{Exact nilpotent nonperturbative BRST symmetry for the
  Gribov-Zwanziger action in the linear covariant gauge}},\ }\href
  {https://doi.org/10.1103/PhysRevD.92.045039} {\bibfield  {journal} {\bibinfo
  {journal} {Phys. Rev. D}\ }\textbf {\bibinfo {volume} {92}},\ \bibinfo
  {pages} {045039} (\bibinfo {year} {2015})},\ \Eprint
  {https://arxiv.org/abs/1506.06995} {arXiv:1506.06995 [hep-th]} \BibitemShut
  {NoStop}%
\bibitem [{\citenamefont {Capri}\ \emph
  {et~al.}(2016{\natexlab{a}})\citenamefont {Capri}, \citenamefont
  {Fiorentini}, \citenamefont {Guimaraes}, \citenamefont {Mintz}, \citenamefont
  {Palhares}, \citenamefont {Sorella}, \citenamefont {Dudal}, \citenamefont
  {Justo}, \citenamefont {Pereira},\ and\ \citenamefont
  {Sobreiro}}]{Capri:2015nzw}%
  \BibitemOpen
  \bibfield  {author} {\bibinfo {author} {\bibfnamefont {M.~A.~L.}\
  \bibnamefont {Capri}}, \bibinfo {author} {\bibfnamefont {D.}~\bibnamefont
  {Fiorentini}}, \bibinfo {author} {\bibfnamefont {M.~S.}\ \bibnamefont
  {Guimaraes}}, \bibinfo {author} {\bibfnamefont {B.~W.}\ \bibnamefont
  {Mintz}}, \bibinfo {author} {\bibfnamefont {L.~F.}\ \bibnamefont {Palhares}},
  \bibinfo {author} {\bibfnamefont {S.~P.}\ \bibnamefont {Sorella}}, \bibinfo
  {author} {\bibfnamefont {D.}~\bibnamefont {Dudal}}, \bibinfo {author}
  {\bibfnamefont {I.~F.}\ \bibnamefont {Justo}}, \bibinfo {author}
  {\bibfnamefont {A.~D.}\ \bibnamefont {Pereira}},\ and\ \bibinfo {author}
  {\bibfnamefont {R.~F.}\ \bibnamefont {Sobreiro}},\ }\bibfield  {title}
  {\bibinfo {title} {{More on the nonperturbative Gribov-Zwanziger quantization
  of linear covariant gauges}},\ }\href
  {https://doi.org/10.1103/PhysRevD.93.065019} {\bibfield  {journal} {\bibinfo
  {journal} {Phys. Rev. D}\ }\textbf {\bibinfo {volume} {93}},\ \bibinfo
  {pages} {065019} (\bibinfo {year} {2016}{\natexlab{a}})},\ \Eprint
  {https://arxiv.org/abs/1512.05833} {arXiv:1512.05833 [hep-th]} \BibitemShut
  {NoStop}%
\bibitem [{\citenamefont {Capri}\ \emph
  {et~al.}(2016{\natexlab{b}})\citenamefont {Capri}, \citenamefont {Dudal},
  \citenamefont {Fiorentini}, \citenamefont {Guimaraes}, \citenamefont {Justo},
  \citenamefont {Pereira}, \citenamefont {Mintz}, \citenamefont {Palhares},
  \citenamefont {Sobreiro},\ and\ \citenamefont {Sorella}}]{Capri:2016aqq}%
  \BibitemOpen
  \bibfield  {author} {\bibinfo {author} {\bibfnamefont {M.~A.~L.}\
  \bibnamefont {Capri}}, \bibinfo {author} {\bibfnamefont {D.}~\bibnamefont
  {Dudal}}, \bibinfo {author} {\bibfnamefont {D.}~\bibnamefont {Fiorentini}},
  \bibinfo {author} {\bibfnamefont {M.~S.}\ \bibnamefont {Guimaraes}}, \bibinfo
  {author} {\bibfnamefont {I.~F.}\ \bibnamefont {Justo}}, \bibinfo {author}
  {\bibfnamefont {A.~D.}\ \bibnamefont {Pereira}}, \bibinfo {author}
  {\bibfnamefont {B.~W.}\ \bibnamefont {Mintz}}, \bibinfo {author}
  {\bibfnamefont {L.~F.}\ \bibnamefont {Palhares}}, \bibinfo {author}
  {\bibfnamefont {R.~F.}\ \bibnamefont {Sobreiro}},\ and\ \bibinfo {author}
  {\bibfnamefont {S.~P.}\ \bibnamefont {Sorella}},\ }\bibfield  {title}
  {\bibinfo {title} {{Local and BRST-invariant Yang-Mills theory within the
  Gribov horizon}},\ }\href {https://doi.org/10.1103/PhysRevD.94.025035}
  {\bibfield  {journal} {\bibinfo  {journal} {Phys. Rev. D}\ }\textbf {\bibinfo
  {volume} {94}},\ \bibinfo {pages} {025035} (\bibinfo {year}
  {2016}{\natexlab{b}})},\ \Eprint {https://arxiv.org/abs/1605.02610}
  {arXiv:1605.02610 [hep-th]} \BibitemShut {NoStop}%
\bibitem [{\citenamefont {Capri}\ \emph
  {et~al.}(2017{\natexlab{a}})\citenamefont {Capri}, \citenamefont {Dudal},
  \citenamefont {Pereira}, \citenamefont {Fiorentini}, \citenamefont
  {Guimaraes}, \citenamefont {Mintz}, \citenamefont {Palhares},\ and\
  \citenamefont {Sorella}}]{Capri:2016gut}%
  \BibitemOpen
  \bibfield  {author} {\bibinfo {author} {\bibfnamefont {M.~A.~L.}\
  \bibnamefont {Capri}}, \bibinfo {author} {\bibfnamefont {D.}~\bibnamefont
  {Dudal}}, \bibinfo {author} {\bibfnamefont {A.~D.}\ \bibnamefont {Pereira}},
  \bibinfo {author} {\bibfnamefont {D.}~\bibnamefont {Fiorentini}}, \bibinfo
  {author} {\bibfnamefont {M.~S.}\ \bibnamefont {Guimaraes}}, \bibinfo {author}
  {\bibfnamefont {B.~W.}\ \bibnamefont {Mintz}}, \bibinfo {author}
  {\bibfnamefont {L.~F.}\ \bibnamefont {Palhares}},\ and\ \bibinfo {author}
  {\bibfnamefont {S.~P.}\ \bibnamefont {Sorella}},\ }\bibfield  {title}
  {\bibinfo {title} {{Nonperturbative aspects of Euclidean Yang-Mills theories
  in linear covariant gauges: Nielsen identities and a BRST-invariant two-point
  correlation function}},\ }\href {https://doi.org/10.1103/PhysRevD.95.045011}
  {\bibfield  {journal} {\bibinfo  {journal} {Phys. Rev. D}\ }\textbf {\bibinfo
  {volume} {95}},\ \bibinfo {pages} {045011} (\bibinfo {year}
  {2017}{\natexlab{a}})},\ \Eprint {https://arxiv.org/abs/1611.10077}
  {arXiv:1611.10077 [hep-th]} \BibitemShut {NoStop}%
\bibitem [{\citenamefont {Capri}\ \emph
  {et~al.}(2017{\natexlab{b}})\citenamefont {Capri}, \citenamefont
  {Fiorentini}, \citenamefont {Pereira},\ and\ \citenamefont
  {Sorella}}]{Capri:2017bfd}%
  \BibitemOpen
  \bibfield  {author} {\bibinfo {author} {\bibfnamefont {M.~A.~L.}\
  \bibnamefont {Capri}}, \bibinfo {author} {\bibfnamefont {D.}~\bibnamefont
  {Fiorentini}}, \bibinfo {author} {\bibfnamefont {A.~D.}\ \bibnamefont
  {Pereira}},\ and\ \bibinfo {author} {\bibfnamefont {S.~P.}\ \bibnamefont
  {Sorella}},\ }\bibfield  {title} {\bibinfo {title} {{Renormalizability of the
  refined Gribov-Zwanziger action in linear covariant gauges}},\ }\href
  {https://doi.org/10.1103/PhysRevD.96.054022} {\bibfield  {journal} {\bibinfo
  {journal} {Phys. Rev. D}\ }\textbf {\bibinfo {volume} {96}},\ \bibinfo
  {pages} {054022} (\bibinfo {year} {2017}{\natexlab{b}})},\ \Eprint
  {https://arxiv.org/abs/1708.01543} {arXiv:1708.01543 [hep-th]} \BibitemShut
  {NoStop}%
\bibitem [{\citenamefont {Capri}\ \emph {et~al.}(2018)\citenamefont {Capri},
  \citenamefont {Dudal}, \citenamefont {Guimaraes}, \citenamefont {Pereira},
  \citenamefont {Mintz}, \citenamefont {Palhares},\ and\ \citenamefont
  {Sorella}}]{Capri:2018ijg}%
  \BibitemOpen
  \bibfield  {author} {\bibinfo {author} {\bibfnamefont {M.~A.~L.}\
  \bibnamefont {Capri}}, \bibinfo {author} {\bibfnamefont {D.}~\bibnamefont
  {Dudal}}, \bibinfo {author} {\bibfnamefont {M.~S.}\ \bibnamefont
  {Guimaraes}}, \bibinfo {author} {\bibfnamefont {A.~D.}\ \bibnamefont
  {Pereira}}, \bibinfo {author} {\bibfnamefont {B.~W.}\ \bibnamefont {Mintz}},
  \bibinfo {author} {\bibfnamefont {L.~F.}\ \bibnamefont {Palhares}},\ and\
  \bibinfo {author} {\bibfnamefont {S.~P.}\ \bibnamefont {Sorella}},\
  }\bibfield  {title} {\bibinfo {title} {{The universal character of
  Zwanziger's horizon function in Euclidean Yang\textendash{}Mills theories}},\
  }\href {https://doi.org/10.1016/j.physletb.2018.03.058} {\bibfield  {journal}
  {\bibinfo  {journal} {Phys. Lett. B}\ }\textbf {\bibinfo {volume} {781}},\
  \bibinfo {pages} {48} (\bibinfo {year} {2018})},\ \Eprint
  {https://arxiv.org/abs/1802.04582} {arXiv:1802.04582 [hep-th]} \BibitemShut
  {NoStop}%
\bibitem [{\citenamefont {{Wolfram Research, Inc.}}()}]{Mathematica}%
  \BibitemOpen
  \bibfield  {author} {\bibinfo {author} {\bibnamefont {{Wolfram Research,
  Inc.}}},\ }\href {https://www.wolfram.com} {\bibinfo {title} {Mathematica
  13.3}}\BibitemShut {NoStop}%
\bibitem [{xAc()}]{xAct1}%
  \BibitemOpen
  \href@noop {} {\bibinfo {title} {{xAct: Efficient tensor computer algebra for
  Mathematica}}},\ \bibinfo {howpublished}
  {\url{http://xact.es/index.html}}\BibitemShut {NoStop}%
\bibitem [{\citenamefont {{Nutma}}(2014)}]{xAct2}%
  \BibitemOpen
  \bibfield  {author} {\bibinfo {author} {\bibfnamefont {T.}~\bibnamefont
  {{Nutma}}},\ }\bibfield  {title} {\bibinfo {title} {{xTras: A field-theory
  inspired xAct package for mathematica}},\ }\href
  {https://doi.org/10.1016/j.cpc.2014.02.006} {\bibfield  {journal} {\bibinfo
  {journal} {Computer Physics Communications}\ }\textbf {\bibinfo {volume}
  {185}},\ \bibinfo {pages} {1719} (\bibinfo {year} {2014})},\ \Eprint
  {https://arxiv.org/abs/1308.3493} {arXiv:1308.3493 [cs.SC]} \BibitemShut
  {NoStop}%
\bibitem [{\citenamefont {Brizuela}\ \emph {et~al.}(2009)\citenamefont
  {Brizuela}, \citenamefont {Martin-Garcia},\ and\ \citenamefont
  {Mena~Marugan}}]{xAct3}%
  \BibitemOpen
  \bibfield  {author} {\bibinfo {author} {\bibfnamefont {D.}~\bibnamefont
  {Brizuela}}, \bibinfo {author} {\bibfnamefont {J.~M.}\ \bibnamefont
  {Martin-Garcia}},\ and\ \bibinfo {author} {\bibfnamefont {G.~A.}\
  \bibnamefont {Mena~Marugan}},\ }\bibfield  {title} {\bibinfo {title} {{xPert:
  Computer algebra for metric perturbation theory}},\ }\href
  {https://doi.org/10.1007/s10714-009-0773-2} {\bibfield  {journal} {\bibinfo
  {journal} {Gen. Rel. Grav.}\ }\textbf {\bibinfo {volume} {41}},\ \bibinfo
  {pages} {2415} (\bibinfo {year} {2009})},\ \Eprint
  {https://arxiv.org/abs/0807.0824} {arXiv:0807.0824 [gr-qc]} \BibitemShut
  {NoStop}%
\bibitem [{\citenamefont {Mertig}\ \emph {et~al.}(1991)\citenamefont {Mertig},
  \citenamefont {Böhm},\ and\ \citenamefont {Denner}}]{FeynCalc1}%
  \BibitemOpen
  \bibfield  {author} {\bibinfo {author} {\bibfnamefont {R.}~\bibnamefont
  {Mertig}}, \bibinfo {author} {\bibfnamefont {M.}~\bibnamefont {Böhm}},\ and\
  \bibinfo {author} {\bibfnamefont {A.}~\bibnamefont {Denner}},\ }\bibfield
  {title} {\bibinfo {title} {Feyn calc - computer-algebraic calculation of
  feynman amplitudes},\ }\href
  {https://doi.org/https://doi.org/10.1016/0010-4655(91)90130-D} {\bibfield
  {journal} {\bibinfo  {journal} {Computer Physics Communications}\ }\textbf
  {\bibinfo {volume} {64}},\ \bibinfo {pages} {345} (\bibinfo {year}
  {1991})}\BibitemShut {NoStop}%
\bibitem [{\citenamefont {Shtabovenko}\ \emph {et~al.}(2016)\citenamefont
  {Shtabovenko}, \citenamefont {Mertig},\ and\ \citenamefont
  {Orellana}}]{FeynCalc2}%
  \BibitemOpen
  \bibfield  {author} {\bibinfo {author} {\bibfnamefont {V.}~\bibnamefont
  {Shtabovenko}}, \bibinfo {author} {\bibfnamefont {R.}~\bibnamefont
  {Mertig}},\ and\ \bibinfo {author} {\bibfnamefont {F.}~\bibnamefont
  {Orellana}},\ }\bibfield  {title} {\bibinfo {title} {{New Developments in
  FeynCalc 9.0}},\ }\href {https://doi.org/10.1016/j.cpc.2016.06.008}
  {\bibfield  {journal} {\bibinfo  {journal} {Comput. Phys. Commun.}\ }\textbf
  {\bibinfo {volume} {207}},\ \bibinfo {pages} {432} (\bibinfo {year}
  {2016})},\ \Eprint {https://arxiv.org/abs/1601.01167} {arXiv:1601.01167
  [hep-ph]} \BibitemShut {NoStop}%
\bibitem [{\citenamefont {Shtabovenko}\ \emph {et~al.}(2020)\citenamefont
  {Shtabovenko}, \citenamefont {Mertig},\ and\ \citenamefont
  {Orellana}}]{FeynCalc3}%
  \BibitemOpen
  \bibfield  {author} {\bibinfo {author} {\bibfnamefont {V.}~\bibnamefont
  {Shtabovenko}}, \bibinfo {author} {\bibfnamefont {R.}~\bibnamefont
  {Mertig}},\ and\ \bibinfo {author} {\bibfnamefont {F.}~\bibnamefont
  {Orellana}},\ }\bibfield  {title} {\bibinfo {title} {{FeynCalc 9.3: New
  features and improvements}},\ }\href
  {https://doi.org/10.1016/j.cpc.2020.107478} {\bibfield  {journal} {\bibinfo
  {journal} {Comput. Phys. Commun.}\ }\textbf {\bibinfo {volume} {256}},\
  \bibinfo {pages} {107478} (\bibinfo {year} {2020})},\ \Eprint
  {https://arxiv.org/abs/2001.04407} {arXiv:2001.04407 [hep-ph]} \BibitemShut
  {NoStop}%
\bibitem [{\citenamefont {Huber}\ and\ \citenamefont {Braun}(2012)}]{DoFun1}%
  \BibitemOpen
  \bibfield  {author} {\bibinfo {author} {\bibfnamefont {M.~Q.}\ \bibnamefont
  {Huber}}\ and\ \bibinfo {author} {\bibfnamefont {J.}~\bibnamefont {Braun}},\
  }\bibfield  {title} {\bibinfo {title} {{Algorithmic derivation of functional
  renormalization group equations and Dyson-Schwinger equations}},\ }\href
  {https://doi.org/10.1016/j.cpc.2012.01.014} {\bibfield  {journal} {\bibinfo
  {journal} {Comput. Phys. Commun.}\ }\textbf {\bibinfo {volume} {183}},\
  \bibinfo {pages} {1290} (\bibinfo {year} {2012})},\ \Eprint
  {https://arxiv.org/abs/1102.5307} {arXiv:1102.5307 [hep-th]} \BibitemShut
  {NoStop}%
\bibitem [{\citenamefont {Huber}\ \emph {et~al.}(2020)\citenamefont {Huber},
  \citenamefont {Cyrol},\ and\ \citenamefont {Pawlowski}}]{DoFun2}%
  \BibitemOpen
  \bibfield  {author} {\bibinfo {author} {\bibfnamefont {M.~Q.}\ \bibnamefont
  {Huber}}, \bibinfo {author} {\bibfnamefont {A.~K.}\ \bibnamefont {Cyrol}},\
  and\ \bibinfo {author} {\bibfnamefont {J.~M.}\ \bibnamefont {Pawlowski}},\
  }\bibfield  {title} {\bibinfo {title} {{DoFun 3.0: Functional equations in
  Mathematica}},\ }\href {https://doi.org/10.1016/j.cpc.2019.107058} {\bibfield
   {journal} {\bibinfo  {journal} {Comput. Phys. Commun.}\ }\textbf {\bibinfo
  {volume} {248}},\ \bibinfo {pages} {107058} (\bibinfo {year} {2020})},\
  \Eprint {https://arxiv.org/abs/1908.02760} {arXiv:1908.02760 [hep-ph]}
  \BibitemShut {NoStop}%
\bibitem [{\citenamefont {Cyrol}\ \emph {et~al.}(2017)\citenamefont {Cyrol},
  \citenamefont {Mitter},\ and\ \citenamefont {Strodthoff}}]{FormTracer}%
  \BibitemOpen
  \bibfield  {author} {\bibinfo {author} {\bibfnamefont {A.~K.}\ \bibnamefont
  {Cyrol}}, \bibinfo {author} {\bibfnamefont {M.}~\bibnamefont {Mitter}},\ and\
  \bibinfo {author} {\bibfnamefont {N.}~\bibnamefont {Strodthoff}},\ }\bibfield
   {title} {\bibinfo {title} {{FormTracer - A Mathematica Tracing Package Using
  FORM}},\ }\href {https://doi.org/10.1016/j.cpc.2017.05.024} {\bibfield
  {journal} {\bibinfo  {journal} {Comput. Phys. Commun.}\ }\textbf {\bibinfo
  {volume} {219}},\ \bibinfo {pages} {346} (\bibinfo {year} {2017})},\ \Eprint
  {https://arxiv.org/abs/1610.09331} {arXiv:1610.09331 [hep-ph]} \BibitemShut
  {NoStop}%
\bibitem [{\citenamefont {Duarte}\ \emph {et~al.}(2016)\citenamefont {Duarte},
  \citenamefont {Oliveira},\ and\ \citenamefont {Silva}}]{Duarte:2016iko}%
  \BibitemOpen
  \bibfield  {author} {\bibinfo {author} {\bibfnamefont {A.~G.}\ \bibnamefont
  {Duarte}}, \bibinfo {author} {\bibfnamefont {O.}~\bibnamefont {Oliveira}},\
  and\ \bibinfo {author} {\bibfnamefont {P.~J.}\ \bibnamefont {Silva}},\
  }\bibfield  {title} {\bibinfo {title} {{Lattice Gluon and Ghost Propagators,
  and the Strong Coupling in Pure SU(3) Yang-Mills Theory: Finite Lattice
  Spacing and Volume Effects}},\ }\href
  {https://doi.org/10.1103/PhysRevD.94.014502} {\bibfield  {journal} {\bibinfo
  {journal} {Phys. Rev. D}\ }\textbf {\bibinfo {volume} {94}},\ \bibinfo
  {pages} {014502} (\bibinfo {year} {2016})},\ \Eprint
  {https://arxiv.org/abs/1605.00594} {arXiv:1605.00594 [hep-lat]} \BibitemShut
  {NoStop}%
\bibitem [{\citenamefont {Cucchieri}\ \emph {et~al.}(2016)\citenamefont
  {Cucchieri}, \citenamefont {Dudal}, \citenamefont {Mendes},\ and\
  \citenamefont {Vandersickel}}]{Cucchieri:2016jwg}%
  \BibitemOpen
  \bibfield  {author} {\bibinfo {author} {\bibfnamefont {A.}~\bibnamefont
  {Cucchieri}}, \bibinfo {author} {\bibfnamefont {D.}~\bibnamefont {Dudal}},
  \bibinfo {author} {\bibfnamefont {T.}~\bibnamefont {Mendes}},\ and\ \bibinfo
  {author} {\bibfnamefont {N.}~\bibnamefont {Vandersickel}},\ }\bibfield
  {title} {\bibinfo {title} {{Modeling the Landau-gauge ghost propagator in 2,
  3, and 4 spacetime dimensions}},\ }\href
  {https://doi.org/10.1103/PhysRevD.93.094513} {\bibfield  {journal} {\bibinfo
  {journal} {Phys. Rev. D}\ }\textbf {\bibinfo {volume} {93}},\ \bibinfo
  {pages} {094513} (\bibinfo {year} {2016})},\ \Eprint
  {https://arxiv.org/abs/1602.01646} {arXiv:1602.01646 [hep-lat]} \BibitemShut
  {NoStop}%
\bibitem [{\citenamefont {Dudal}\ \emph {et~al.}(2012)\citenamefont {Dudal},
  \citenamefont {Oliveira},\ and\ \citenamefont
  {Rodriguez-Quintero}}]{Dudal:2012zx}%
  \BibitemOpen
  \bibfield  {author} {\bibinfo {author} {\bibfnamefont {D.}~\bibnamefont
  {Dudal}}, \bibinfo {author} {\bibfnamefont {O.}~\bibnamefont {Oliveira}},\
  and\ \bibinfo {author} {\bibfnamefont {J.}~\bibnamefont
  {Rodriguez-Quintero}},\ }\bibfield  {title} {\bibinfo {title} {{Nontrivial
  ghost-gluon vertex and the match of RGZ, DSE and lattice Yang-Mills
  propagators}},\ }\href {https://doi.org/10.1103/PhysRevD.86.105005}
  {\bibfield  {journal} {\bibinfo  {journal} {Phys. Rev. D}\ }\textbf {\bibinfo
  {volume} {86}},\ \bibinfo {pages} {105005} (\bibinfo {year} {2012})},\
  \Eprint {https://arxiv.org/abs/1207.5118} {arXiv:1207.5118 [hep-ph]}
  \BibitemShut {NoStop}%
\end{thebibliography}%

\end{document}